\documentclass[useAMS,usenatbib]{mn2e}
\usepackage{epsfig,amsmath,amssymb,amsfonts,mathrsfs,latexsym,graphicx}

\usepackage[draft]{hyperref}
\usepackage{natbib}
\usepackage{color}
\usepackage{algorithm}
\usepackage{ textcomp } 

\usepackage{fixltx2e}

\usepackage{float}
\usepackage[caption = false]{subfig}
\usepackage{multirow}
\def\apjl{ApJ}%
\def\apjs{ApJS}%
\def\ao{Appl.~Opt.}%
\def\aap{A\&A}%
\graphicspath{{images/}}

\newcommand{\kmsMpc}{\,km\,s$^{-1}$Mpc$^{-1}$}

\newcommand{\Msun}{{\rm M}_{\odot}}

\newcommand{\kms}{km~s$^{-1}$}

\newcommand{\SiII}{Si~{\sc ii}}

\newcommand{\CaII}{Ca~{\sc ii}}

\newcommand{\Cofs}{$^{56}$Co}
\newcommand{\Nifs}{$^{56}$Ni}

\newcommand{\Deltam}{$\Delta m_{15}(B)$}

\def\gsim{\mathrel{\rlap{\lower 4pt \hbox{\hskip 1pt $\sim$}}\raise 1pt \hbox {$>$}}}
\def\lsim{\mathrel{\rlap{\lower 4pt \hbox{\hskip 1pt $\sim$}}\raise 1pt \hbox {$<$}}}

\title[Type Ia supernova spectra and models]{A metric space for type Ia
supernova spectra: a new method to assess explosion scenarios}

\author[M. Sasdelli et al.]
{\parbox{\textwidth}{\vspace{-.5cm} \large  Michele~Sasdelli$^{1}$\thanks{E-mail: 
sasdelli@mpa-garching.mpg.de}, 
 W.~Hillebrandt$^{1}$,
M.~Kromer$^{2,4,5}$,
E.~E.~O.~Ishida$^{3,1}$,
F.~K.~R\"{o}pke$^{4,5}$,
S.~A.~Sim$^{6,7}$,
R.~Pakmor$^{4}$,
I.~R.~Seitenzahl$^{8,7}$,
M.~Fink$^{9}$
}\vspace{0.6cm}\\ 
\parbox{\textwidth}{ 
$^{1}$Max-Planck-Institut f\"ur Astrophysik, Karl-Schwarzschild-Str. 1, 85741 Garching bei M\"unchen, Germany\\
$^{2}$The Oskar Klein Centre \& Dept. of Astronomy, Stockholm University, AlbaNova, SE-106 91 Stockholm, Sweden\\
$^{3}$Clermont Universit\'e, Universit\'e Blaise Pascal, CNRS/IN2P3, Laboratoire de Physique Corpusculaire, BP 10448, F-63000 \\ Clermont-Ferrand, France\\
$^{4}$Heidelberger Institut f\"{u}r Theoretische Studien, Schloss-Wolfsbrunnenweg 35, 69118 Heidelberg, Germany\\
$^{5}$Zentrum f\"ur Astronomie der Universit\"at Heidelberg, Institut f\"ur Theoretische Astrophysik, Philosophenweg 12, 69120 Heidelberg, Germany\\
$^{6}$Astrophysics Research Centre, School of Mathematics and Physics, Queen's University Belfast, Belfast BT7 1NN, UK\\
$^{7}$ARC Centre of Excellence for All-sky Astrophysics (CAASTRO)\\
$^{8}$Research School of Astronomy and Astrophysics, Australian National University, Canberra, ACT 2611, Australia\\
$^{9}$ Institut f\"{u}r Theoretische Physik und Astrophysik, Universit\"{a}t W\"{u}rzburg, Emil-Fischer-Str. 31, 97074 W\"{u}rzburg, Germany\\
}}

\begin{document}

\date{Accepted 2016 December 19; Received 2016 November 28; in original form 2016 April 2}

\pagerange{\pageref{firstpage}--\pageref{lastpage}} \pubyear{2017}

\maketitle
\label{firstpage}

\begin{abstract}

Over the past years type Ia supernovae (SNe\,Ia) have become a
major tool to determine the expansion history of the Universe, and
considerable attention has been given to, both, observations and models
of these events. However, until now, their progenitors are not known.
 The observed diversity of light curves and spectra seems to
point at different progenitor channels and explosion mechanisms.
Here, we present a new way to compare model predictions with observations
in a systematic way. Our method is based on the construction of a metric
space for SN\,Ia spectra by means of linear Principal Component
Analysis (PCA), taking care of missing and/or noisy data, and making use
of Partial Least Square regression (PLS) to find correlations between
spectral properties and photometric data. We investigate realizations of
the three major classes of explosion models that are presently
discussed: delayed-detonation Chandrasekhar-mass explosions,
sub-Chandrasekhar-mass detonations, and double-degenerate mergers, and
compare them with data. We show that in the PC space all scenarios have
observed counterparts, supporting the idea that different progenitors are
likely. However, all classes of models face problems in reproducing the
observed correlations between spectral properties and light curves and
colors. Possible reasons are briefly discussed.

\end{abstract}


\section{Introduction}
\label{sec:intro}

{\color{black}Type Ia supernovae (SN\,Ia) are thought to be the
thermonuclear explosion of a white dwarf in a binary
system \citep{1960ApJ...132..565H}.  It is debated whether
the companion is a second white
dwarf \citep{1984ApJS...54..335I,1984ApJ...277..355W} or a
non-degenerate star \citep{1973ApJ...186.1007W}.  A direct detection
of the progenitor is still missing, this is why} to construct models
from first principles is currently a promising strategy to constrain
the progenitors and explosion mechanisms of SNe\,Ia
\citep{2000ARA&A..38..191H}.

In doing so, one assumes a progenitor system and explosion scenario and
simulates nuclear burning and the explosion in detail. 
{\color{black}The comparison to the observables requires radiation
 transport simulations.} Currently
investigated models are described in detail in a recent review by
\cite{2013FrPhy...8..116H}. By varying the (physical) input parameters
different realizations of every scenario are obtained.  Since the
progenitors of SNe\,Ia are not known, in most cases the various scenarios
are simulated for a wide range of reasonable parameter values in order
to see if the observed diversity of SNe\,Ia can be explained
\citep{2013MNRAS.429.1156S,2014MNRAS.438.1762F}.
 For example, varying the initial mass in case of sub-Chandrasekhar
mass double-detonation models changes the predicted luminosity of the
explosion
\citep{2010ApJ...719.1067K,2010ApJ...714L..52S,2011ApJ...734...38W,2013ApJ...774..137M}.
Alternatively, in some cases specific realizations were studied as
possible explanations of unusual events
\citep[e.g.][]{2010Natur.463...61P,2013ApJ...778L..18K}.

To be more specific, one computes synthetic light curves and time
sequences of spectra for the models and compares them with
observations. The production of the light from the radioactive decay
of \Nifs\ and \Cofs\ is calculated and the propagation of photons and
their escape from the ejecta is computed, in 3-dimensions usually by
means of Monte-Carlo
methods \citep{2006ApJ...651..366K,2009MNRAS.398.1809K}. However, this
approach is computationally expensive and so far only a small part of
the parameter space of explosion scenarios was investigated.
Moreover, it allows no freedom to adjust the resulting synthetic
observables to fit the observations.  Finally, it is not easy to use
observations to guide the explosion modeling.  {\color{black}By
``standard methods'' it is possible to compare individual models to
individual SNe, but it is particularly difficult to compare a group of
models with a population of supernovae in a systematic and
quantitative way.}

The current approach to test models against observations is mostly done by
comparing the light curves of groups of models with the known global
properties of light curves of SN\,Ia. One of the most important
and best known of these properties is the Phillips relation, that is,
the correlation between the decline rate of the light curve and the
luminosity at peak \citep{1993ApJ...413L.105P}. Due to the relative
simplicity of light curves of SNe\,Ia it is easy to assess whether
realizations of a given explosion scenario follow the Phillips relation
or not. However, comparing the global properties of spectra with model
predictions is a much harder challenge.  Up to now it is done mostly
qualitatively ('$\chi^2$ by eye') on a case-by-case basis, that is, by
comparing a specific model from a given scenario with an individual
supernova \citep[or a representative example of a particular SN class,
e.g.][]{2012ApJ...750L..19R}. It is obvious, that within this
approach models cannot be tested against empirical relations between
different spectral properties and between spectra and light curves of
real supernovae, which would be more constraining for the models.

Here we are using a different approach. In a first step we construct a
'metric space' for SN\,Ia spectral time series by means of a {\sl
Principal Component Analysis} (PCA) based on a large sample of observed
SNe\,Ia \citep[see][and Section~\ref{subsec:pca_space}, respectively, 
for a description of the method and of the database]{2015MNRAS.447.1247S}.
Using PCA it is possible to discover correlations
between spectral properties (if they exist) and empirical relations
between spectra and photometry can be studied systematically with {\sl
Partial Least Square} regression (PLS, Section \ref{subsec:PLS_analysis}). 
Next, the principal components of synthetic
spectra of models can be computed and placed into the PCA-space of the
data. {\color{black}The consistency between spectral properties and 
broad-band photometry of the models can be checked with PLS regression.}
 It will be shown that this {\color{black}combined} approach allows us to derive
constraints for the models in a more systematic way than was previously
possible.

The present paper is organized as follows. In Section~2 we summarize the
essentials of Expectation Maximization {\sl PCA} and {\sl PLS} as developed in
\cite{2015MNRAS.447.1247S}. In Section 3 we discuss briefly the
explosion models that we compare with the data in Section 4.  A
summary and conclusions are presented in Section 5.

\section{Metric space on public SN\,Ia spectra}

In this paper we make use of a set of techniques developed
 in \cite{2015MNRAS.447.1247S} for the study of SN\,Ia spectral time
 series and photometry. Our previous analysis was based on data from
 the Nearby Supernova
 Factory \citep[SNf,][]{2002SPIE.4836...61A}. This is a collection of
 spectrophotometric time series of a large sample of mostly normal
 SNe\,Ia. Since one of the main goals of the SNf survey is to
 construct a SN\,Ia Hubble diagram, for the most part supernovae were
 followed only if they were normal SNe\,Ia in the smooth Hubble flow.

\subsection{The data set}

For this study we decided to apply the method
of \cite{2015MNRAS.447.1247S} to spectroscopic datasets from the
literature. As this method is largely insensitive to errors in flux
calibration, it allows the use of a larger fraction of SNe, in
particular those very nearby supernovae with well observed low noise
spectra. These data generally include larger fractions of peculiar SNe
since peculiarity is often the impetus for obtaining a good series of
spectroscopic observations. Since we wish to use the technique
of \cite{2015MNRAS.447.1247S} to explore which models map to {\it any}
observed SN, such peculiar SNe are valuable for comparison with
models.

Finally, these data offer a larger number of well-observed early
spectra (earlier than a week before maximum light in $B$-band). The
early behaviour is crucial to constrain models and, in addition, the
approximations used in the radiation transport code are more reliable
at early times. Therefore, in order to construct a metric space for
our analysis of model properties we collected a large sample of SN\,Ia
spectra available in the literature. The sources are the CfA
spectroscopic release \citep{2012AJ....143..126B}, the Berkeley
Supernova Program
\citep{2012MNRAS.425.1789S}, the Carnegie Supernova Project
\citep[CSP, ][]{2013ApJ...773...53F}.  We also use SN catalogs such as SUSPECT
\footnote{\url{http://www.nhn.ou.edu/~suspect}} and WISEREP
\citep{2012PASP..124..668Y}. The spectra are de-redshifted by using the
heliocentric redshifts tabulated in \cite{2012AJ....143..126B}. CSP spectra are
published in rest frame.

As representative subsets of these spectra have been shown to be
insufficiently spectrophotometric \citep{2008AJ....135.1598M,
2012AJ....143..126B, 2012MNRAS.425.1789S, 2013ApJ...773...53F}, for
some of the studies in this analysis, separate photometry from broadband
imaging is needed.
The photometry and the $B-V$ colors are collected from
\cite{2009ApJ...700..331H}.  They were obtained from light-curve fitting
using MLCS2k2 \citep{2007ApJ...659..122J}. The photometry and colors were
K-corrected, corrected for Milky Way extinction, and corrected for time
dilation.  The host-galaxy extinction was not removed. The CSP photometry
comes from
\cite{2011AJ....142..156S}.

The $B$-band photometry is transformed into absolute magnitudes using
the CMB centered redshift measurements from \cite{2009ApJ...700..331H}.
The error of the absolute magnitude is computed adding in quadrature an
error due to the peculiar motion of the host galaxy. We assume a
standard deviation of 500 \kms\ for this peculiar velocity
\citep{2003MNRAS.346...78H}.

The training set we used finally for the analysis consisted of
238 supernovae and a total of 2154 spectra. They were binned according
to Table 1. Not all of them were spectroscopically classified as in 
\cite{2012AJ....143..126B}, but the latter had 112 ``normal'', 20
``91T-like'', 16 ``91bg-like'', 34 ``high-velocity photospheric \SiII\"~,
and 6 ``peculiar'' supernovae in. These fractions of different
subtypes appear to be typical for the sample we use (see also
Figs. \ref{fig:scatter_plot_sne} and \ref{fig:scatter_plot_sne_2}).

A few typical examples of data we used in our analysis are given in
Appendix A.

\begin{table}{}
\caption{Binning and number of spectra. Time is measured in days
relative to B-band maximum.}
\begin{center}
\begin{tabular}{lll}
\label{tab:spectra}
t$_{min}$ & t$_{max}$ & Number \\
 &  &   of spectra   \\
\hline  
  $-12.5$ & $-10.0$ & ~90 \\
  $-10.0$ & ~$-7.5$  & 165 \\
  ~$-7.5$ & ~$-5.0$  & 218 \\
   ~$-5.0$ & ~$-2.5$  & 233 \\
   ~$-2.5$ &  ~~~0.0  & 256 \\
   ~~~0.0 &  ~~~2.5  & 258 \\
    ~~~2.5 &  ~~~5.0  & 231 \\
    ~~~5.0 &  ~~~7.5  & 177 \\
    ~~~7.5 & ~~10.0  & 170 \\
   ~~10.0 & ~~12.5  & 160 \\
   ~~12.5 & ~~15.0  & 101 \\
   ~~15.0 & ~~17.5  & ~95 \\    
\hline
\end{tabular}
\end{center}
\end{table}

\subsection{The method}

In general terms, Principal Component Analysis (PCA) is a standard
statistical tool for data reduction. It reduces the dimensionality of
a data set with an intrinsically high number of dimensions, but
retaining the meaningful information. PCA is essentially a rotation of
the axes of the high dimensional space representing the data that aligns
the first axis to the direction with the largest variance, the 1st
principal component (PC). The second PC maximizes the variance,
subject to being orthogonal to the 1st, and so on. Thus, an initially
multivariate data set is described by a smaller number of uncorrelated
parameters.

Here we present a brief summary of a variant of the method developed
in \cite{2015MNRAS.447.1247S} and refer for details to our
previous work. In order to study systematically the spectral
characteristics of different explosion scenarios we use a method to
construct a {\it metric space} for SN\,Ia spectral time series. We use
an {\sl Expectation Maximization Principal Component
Analysis} \citep[EMPCA][]{roweis1998algorithms, 2012PASP..124.1015B}
on the space of spectral series. The most important idea behind our
approach is to study the derivative of the flux over wavelength
$(\partial F(\lambda)/\partial \lambda)$ of the spectra instead of the
flux $(F(\lambda))$ itself.
The use of derivative spectroscopy suppresses the variance associated
with reddening, uncertainties in the distance and calibration of the
flux. Another key ingredient is the use of series of epochs to include
the information of spectral evolution \citep[e.g. velocity
gradients,][]{2005ApJ...623.1011B}, and not only 
spectral indicators at maximum. The final step is the use of {\sl Partial Least
Square} regression (PLS). This is a robust way to find relations
between spectra and photometric properties, such as absolute
magnitudes and intrinsic colors. The method to construct the {\it
metric space} and apply PLS regression is detailed
in \cite{2015MNRAS.447.1247S}. In this section we describe again the
crucial steps of the method and explain a few incremental improvements.

\subsubsection{Derivative spectroscopy}

In order to obtain meaningful derivatives of the fluxes the observed spectra
have to be smoothed. Here we apply an improved Savitzky-Golay
filter \citep{1964AnaCh..36.1627S} for smoothing. A well known way to improve
low band pass filters is to iterate the filtering a number of
times \citep{kaiser1977sharpening}.  {\color{black}The fit of a third grade
polynomial on a window of 1800 \kms\ is iterated 5 times.}  With this
filtering we improved the rejection of the noise and at the same time obtained
sharper spectral features to be fed into the PCA algorithm.

\subsubsection{Spectral series and EMPCA}

\label{subsec:pca_space}


\begin{figure*}
\centering
\subfloat{\includegraphics[width=2.\columnwidth]{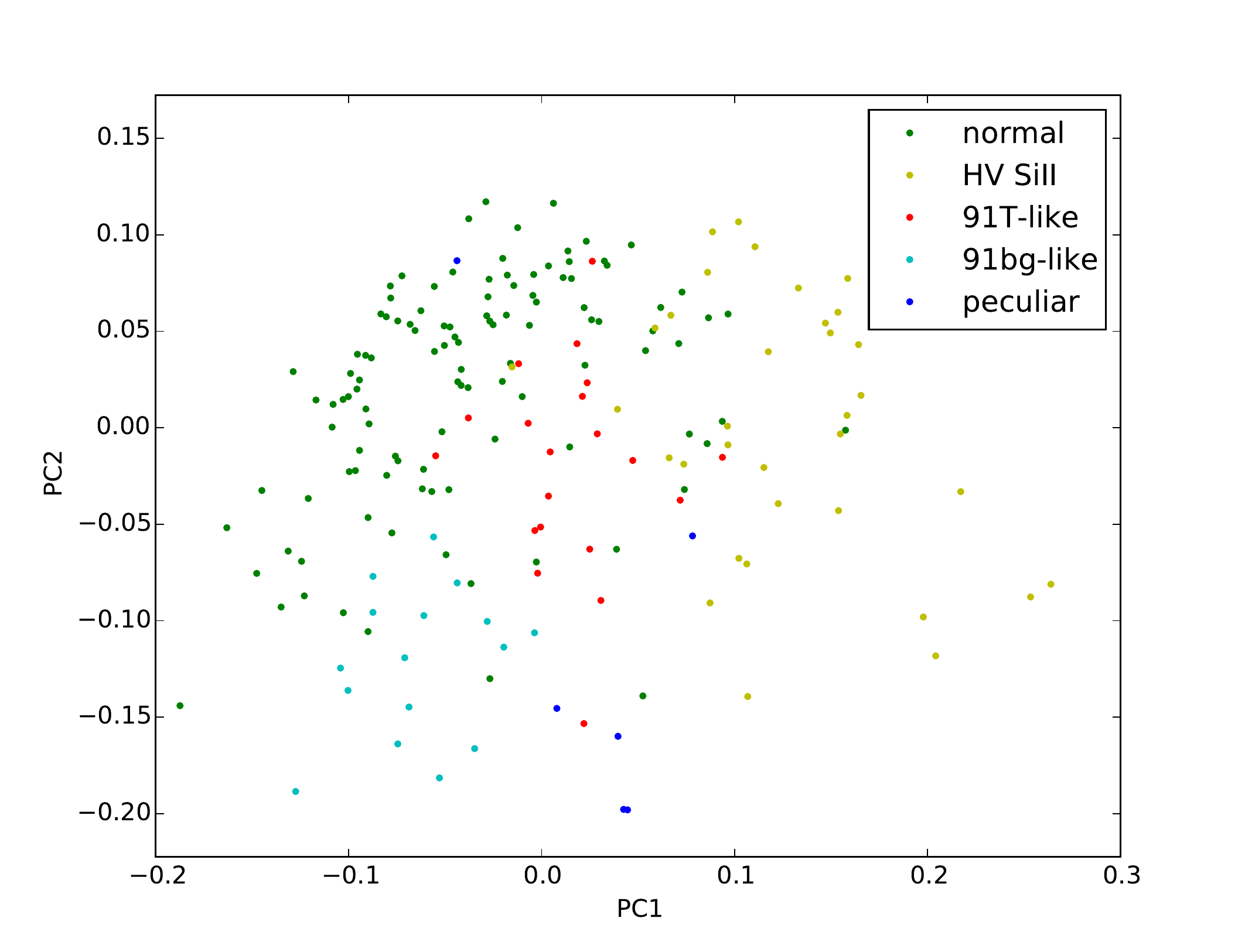}}
\caption{Plot of the first two principal components for the SNe\,Ia of
our sample. 
The classification is according to \protect\cite{2012AJ....143..126B}.}
\label{fig:scatter_plot_sne} \end{figure*}

\begin{figure*}
\centering
\subfloat{\includegraphics[width=2.\columnwidth]{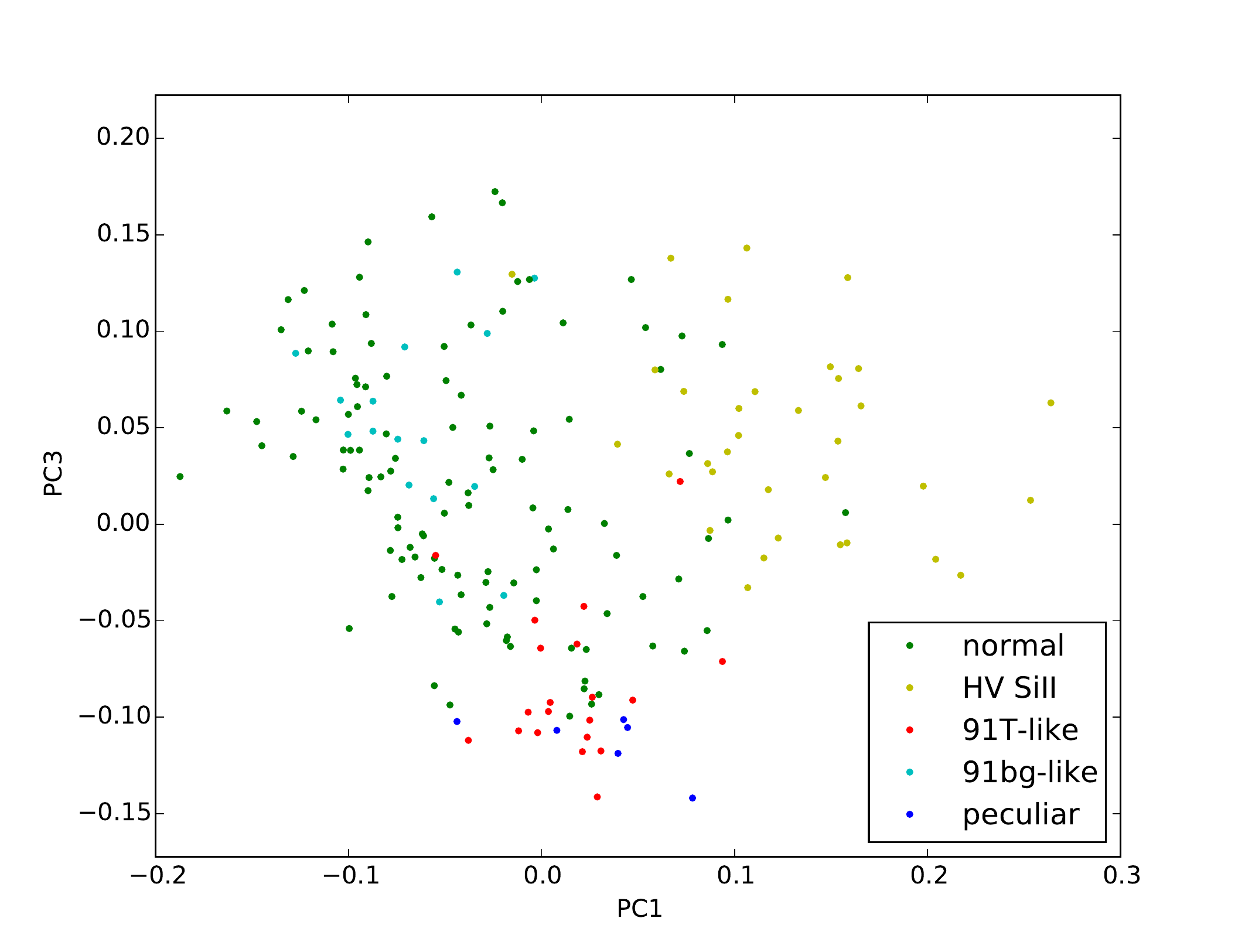}}
\caption{ Plot of the first and the third principal components for the
SNe\,Ia of our sample. This is a viewing angle on our metric space
different from Fig. \ref{fig:scatter_plot_sne}.  The classification is
according to \protect\cite{2012AJ....143..126B}.     }
\label{fig:scatter_plot_sne_2} 
\end{figure*}

The input matrix for the analysis is {\color{black}formed by
 observations (rows) and observables (columns).  Every supernova is an
 individual observation and the derivatives of the fluxes of spectra
 at different epochs are treated as different observables.} The input
 vectors of this matrix are constructed concatenating spectra at
 different epochs in the spectral series.  The EMPCA code
 of \cite{2012PASP..124.1015B} deals automatically with missing epochs
 and/or wavelength gaps in the data.  The derivative analysis
 frees us from the need of flux calibrated spectra with known
 distances, and the large number of low noise spectra for nearby
 SNe allows us to increase the included range of epochs over what was
 done in \cite{2015MNRAS.447.1247S}. We now include spectra from
 $-12.5$ days up to $+17.5$ days from $B$-maximum.  The spectral
 coverage of CfA supernovae, by number the largest of the sample, is
 usually limited to below $\sim 7000$\AA.  This restricts our current
 analysis to a spectral range between $\sim3500$\AA\ and
 $\sim7000$\AA.  This is not a major issue here, since some of
 the information red-ward of $7000$\AA\ (mostly in the IR triplet of
\CaII) is also present in the included wavelength
 range as was shown in \cite{2015MNRAS.447.1247S}.  
The {\it metric space} resulting from PCA has a low dimensionality. The output
consists of just five significant components.

The metric space obtained from public data is similar to the one
obtained in our previous work on SNf data. Projecting the supernovae
on the first three principal components
(Figs.  \ref{fig:scatter_plot_sne} and \ref{fig:scatter_plot_sne_2})
shows the groups found in our previous work also \citep[Fig.  7
of][]{2015MNRAS.447.1247S}.  Using the first $4 - 5$ dimensions, the
metric space obtained from the public SNe\,Ia clearly distinguishes
the spectroscopic subtypes.  Normal SNe\,Ia are on the top-left side
of the cloud of points in Fig.
\ref{fig:scatter_plot_sne}, supernovae with a high velocity photospheric
\SiII~6355~\AA\ have a large first component, 1991T-like events have
negative third component (Fig. \ref{fig:scatter_plot_sne_2}). In the
public sample we have also fainter supernovae.  There is a significant
number of 1991bg-like SNe, characterized by fast declining light
curves, low luminosity and low temperature of the spectra. At the
bottom of Fig.
\ref{fig:scatter_plot_sne}, further apart than 1991T-like, there are a
number of supernovae called peculiar by \cite{2012AJ....143..126B}.
Many of them are 2002cx-like.  This is another class of faint
objects characterized by hot spectra and very low line velocities. These
faint SNe\,Ia are absent in the SNf cosmology sample and are
 not well represented in the publicly available data either.

In Appendix A we present several examples showing the quality of
our PCA reconstruction as well as possible problems for supernovae of
various sub-types. We demonstrate that the method works fine even
in case of noisy spectra and/or missing data, provided there is a
sufficient number of similar objects in the training set.

\subsubsection{PLS regression}
\label{subsec:PLS_analysis}

The final and crucial ingredient in our approach is a {\sl Partial Least
Square} regression
\citep[PLS,][]{{wold1982soft},{wold1984collinearity}}.  It is also known
as {\sl Projection to Latent Structures}, the latter name describing
better the aim of the technique. With this tool we find the empirical
relations between the spectral properties encoded in the PCA space and
photometric properties, such as \Deltam, the intrinsic part of the $B-V$
color, or the absolute $B$ magnitude. The underlying assumption is that the
intrinsic color (absolute $B$ magnitude or \Deltam) is a function of the
spectral properties of the supernova and can be predicted from them.

With this {\it metric space} we can check if different explosion
scenarios are consistent with empirical relationships between spectral
and photometric properties that hold for SNe\,Ia and thereby analyse
potential advantages or shortcomings of the various classes of models.

\section{Explosion scenarios}

It is widely accepted that SNe\,Ia are the result of the thermonuclear
explosion of a carbon-oxygen white dwarf triggered by the
interaction with a companion star. But beyond this very little is
known with certainty. In this section we briefly review
the presently favored scenarios (see also \citealt{2013FrPhy...8..116H}
for a recent review).

\subsection{Delayed detonations}
\label{sec:deldet}

The delayed detonation model \citep{1991A&A...245..114K} is one of the
most studied scenarios as an explanation for SNe\,Ia.  This explosion
mechanism is usually proposed for single-degenerate systems where a
carbon-oxygen white dwarf explodes close to the Chandrasekhar mass
M$_{\text{Chan}}$ ($\sim1.4 \Msun$) after accreting mass from a non-degenerate
companion, presumably through Roche-lobe overflow. The matter steadily
burns to carbon and oxygen on the surface of the white dwarf
increasing its mass until the density at the center is sufficient for
the ignition of nuclear burning and for a combustion wave to form.

In this class of models it is assumed that in the beginning burning
proceeds with a flame speed lower than the speed of sound (deflagration)
and incinerates the interior of the star. This phase allows for the
white dwarf to expand and decrease the density of the unburned material,
a necessary ingredient for the synthesis of intermediate mass elements
(IME). It is further assumed that in a next step a transition from a
deflagration to a detonation takes place somewhere in the star with a
burning velocity now larger than the speed of sound.  Whether or not this
happens in reality is heavily disputed
\citep{2007ApJ...668.1109W,2007ApJ...668.1103R,2010ApJ...710.1654A,
2010ApJ...710.1683S,2011PhRvL.107e4501P,2013A&A...550A.105C}, but
it cures several of the problems of pure-deflagration models (see
 Section~\ref{sec:pure_def}), i.e.,
these models can be brighter and have less unburned carbon and oxygen at
low velocity. In fact, after the transition the detonation front quickly
burns most of the remaining fuel, partially to \Nifs\ and partially to
IMEs
\citep{1996ApJ...457..500H,2007Sci...315..825M,2009Natur.460..869K,
2011MNRAS.417.1280B,2013MNRAS.429.2127B,2013MNRAS.436..333S}.
It is this property of the delayed detonation models that brings
Chandrasekhar-mass explosions closer to the observed light curves and
spectra of normal SN\,Ia than pure-deflagration models.

\subsection{Pure deflagrations}
\label{sec:pure_def}

Historically, pure deflagrations of M$_{\text{Chan}}$ Carbon-Oxygen
white dwarfs such as the W7 model \citep{1984ApJ...286..644N} have
been favoured as an explanation for normal SNe\,Ia.  The W7 model is a
1D deflagration with a parametrized flame speed, assumed to be
proportional to the distance from the center of the star.  However, in
this model the burning proceeds faster than what happens in recent 3D
deflagrations \citep{2007ApJ...668.1132R,2014MNRAS.438.1762F,2014ApJ...782...11M,2014ApJ...789..103L}
leading to nuclear burning comparable to the delayed detonation
models.  W7 is generally in reasonable agreement with normal
SNe\,Ia, although at later times it seems to have a core
structure different from the prototypical normal SN 1994D
\citep{2001ApJ...557..266L}.

In recent deflagration models the initial
conditions are the same as in the delayed detonations, but the (unproven)
deflagration to detonation transition (see Section~\ref{sec:deldet}) does not
happen. Consequently, for
equivalent initial conditions, the burning is less complete than in the
corresponding delayed detonation models. The production of \Nifs\ is more limited and
this limits the maximum possible luminosity of the scenario
\citep{2007ApJ...668.1132R}. 
Hence this mechanism cannot be an explanation for the brightest SNe\,Ia
but may explain some faint peculiar subtypes of SNe\,Ia  
\citep{2012ApJ...761L..23J,2013MNRAS.429.2287K,2015MNRAS.450.3045K,2015A&A...573A...2S}.

\subsection{Sub-Chandrasekhar mass detonations}

The Sub-Chandrasekhar mass models investigated here, i.e., exploding white
dwarfs with a mass lower than M$_{\text{Chan}}$, are the result of detonations
ignited near the center of the white dwarf.
The rates of this progenitor channel are easier to explain using binary population
synthesis simulations \citep{2011MNRAS.417..408R} than the single-degenerate
scenario.
 In contrast to the M$_{\text{Chan}}$ case,
for these stars the density at the center is not high enough to self-ignite
carbon and oxygen but a trigger is needed. A possible mechanism is the
so-called double-detonation \citep{1990ApJ...354L..53L}. A layer of
helium-rich material on the surface of the white dwarf may detonate first, for
example after it was accreted from a companion (He-)star or during a merger
with the companion.  The He-detonation will engulf the white dwarf sending
shock waves inward which will converge close to the center. Numerical
simulations have
shown \citep{2007A&A...476.1133F,2010A&A...514A..53F,2013ApJ...774..137M} that
in the converging shocks the temperature increases sufficiently to trigger a
secondary detonation in the C+O fuel.  Burning at the lower density of the
sub-M$_{\text{Chan}}$ white dwarf (as compared to the M$_{\text{Chan}}$ case)
produces naturally a large amount of IME (that are seen in the ejecta), and
the mass of the initial white dwarf (setting its density) is an excellent
parameter to drive the mass of \Nifs\ and reproduce the variance in luminosity
observed in SNe\,Ia \citep{2010ApJ...714L..52S}.  We note,
however, that the models of \cite{2010ApJ...714L..52S} are explosions of bare
CO-cores which may be more generic but less realistic than specific double
detonations including the He shell.

\subsection{Double-degenerate mergers}

A scenario different from the previous ones is the merger
of two sub-M$_{\text{Chan}}$ C+O white dwarfs \citep{1984ApJS...54..335I,1984ApJ...277..355W}.
The comparison of recent population synthesis simulations
\citep[e.g.][]{2009ApJ...699.2026R,2012A&A...546A..70T} with the observed rates favours 
 this scenario over single-degenerate progenitors.
 The orbit of such binary systems slowly
decays through gravitational wave emission until the two stars 
merge.  If this happens on a timescale shorter than the Hubble time,
the binary may be a candidate for a SN\,Ia. {\color{black}The process of merging
may trigger an explosion in the primary through the double-detonation mechanism
explained in the previous section \citep{2013ApJ...770L...8P}.  If an adequate
He layer is not present, the rapid accretion due to tidal
interaction may form a hot spot on the primary with a high enough
density to trigger the detonation of carbon there
\citep{2012ApJ...747L..10P,2014ApJ...785..105M}. 

If, however, the burning does not start promptly, the secondary
will be disrupted over a few orbits and will be accreted onto the
primary on a secular time scale.
If during this process the mass of the primary gets close to
M$_{\text{Chan}}$ either an explosion is triggered at the center or the white
dwarf may collapse to a neutron star \citep{1980PASJ...32..303M,1985A&A...150L..21S}. In case of an explosion this scenario
might resemble fast rotating M$_{\text{Chan}}$ models.

Here, we will investigate models only in which the explosion is
triggered promptly. As in the case of sub-M$_{\text{Chan}}$
detonations, the mass of \Nifs\ is largely determined by the mass of the
primary white-dwarf. The mass of the secondary and the viewing angle
could be additional parameters explaining the diversity within SNe\,Ia.
Alternatively, if the two white dwarfs have comparable masses both
stars are disrupted during the interaction.
In the case of two $0.9\Msun$ white dwarfs it was found that the low central
density leads
to little \Nifs\ production and to a significantly lower luminosity compared
to mergers with a similar but unequal and larger progenitor masses.
Therefore, it was suggested as a scenario for the SN~1991bg-like subluminous
supernovae.

\section{The models in PCA space} 
\label{sec:PCAspace}

\begin{figure*}
\centering
\subfloat{\includegraphics[width=2.\columnwidth]{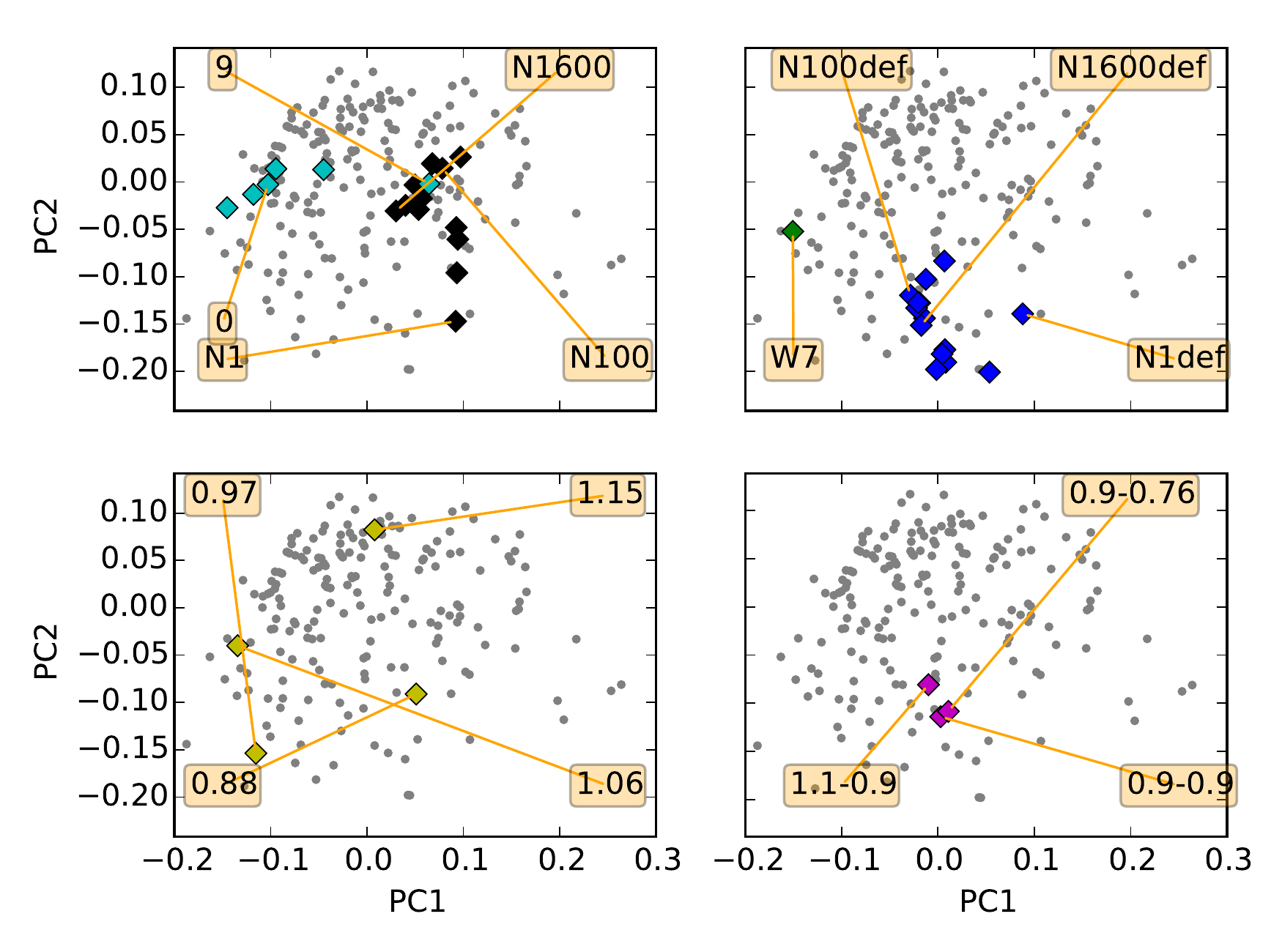}}
\caption{ The first two principal components of the data (grey dots) with the
projections of the analyzed models over-plotted. Most of the models are well
inside the PC space determined by the data {\color{black}in the most important
components}.  Model series are characterized by chains of points in a
multidimensional space.  The top-right
diagram shows the position of W7 on the low luminous edge of normal SN\,Ia and
the 3D deflagrations populating the space of the faint 02cx-like SNe. The
top-left diagram shows the series delayed detonation models with a variable
number of ignition spots (black) and models with the composition of N100 and a
progressively larger degree of mixing ($0$ to $9$, cyan).  The bottom-left
diagram shows the sub-Chandra detonations with different initial masses
(yellow). The bottom-right panel shows three merger models with different
initial masses (magenta). }
\label{fig:scatter_plot}
\end{figure*}

\begin{figure*}
\centering
\subfloat{\includegraphics[width=2.\columnwidth]{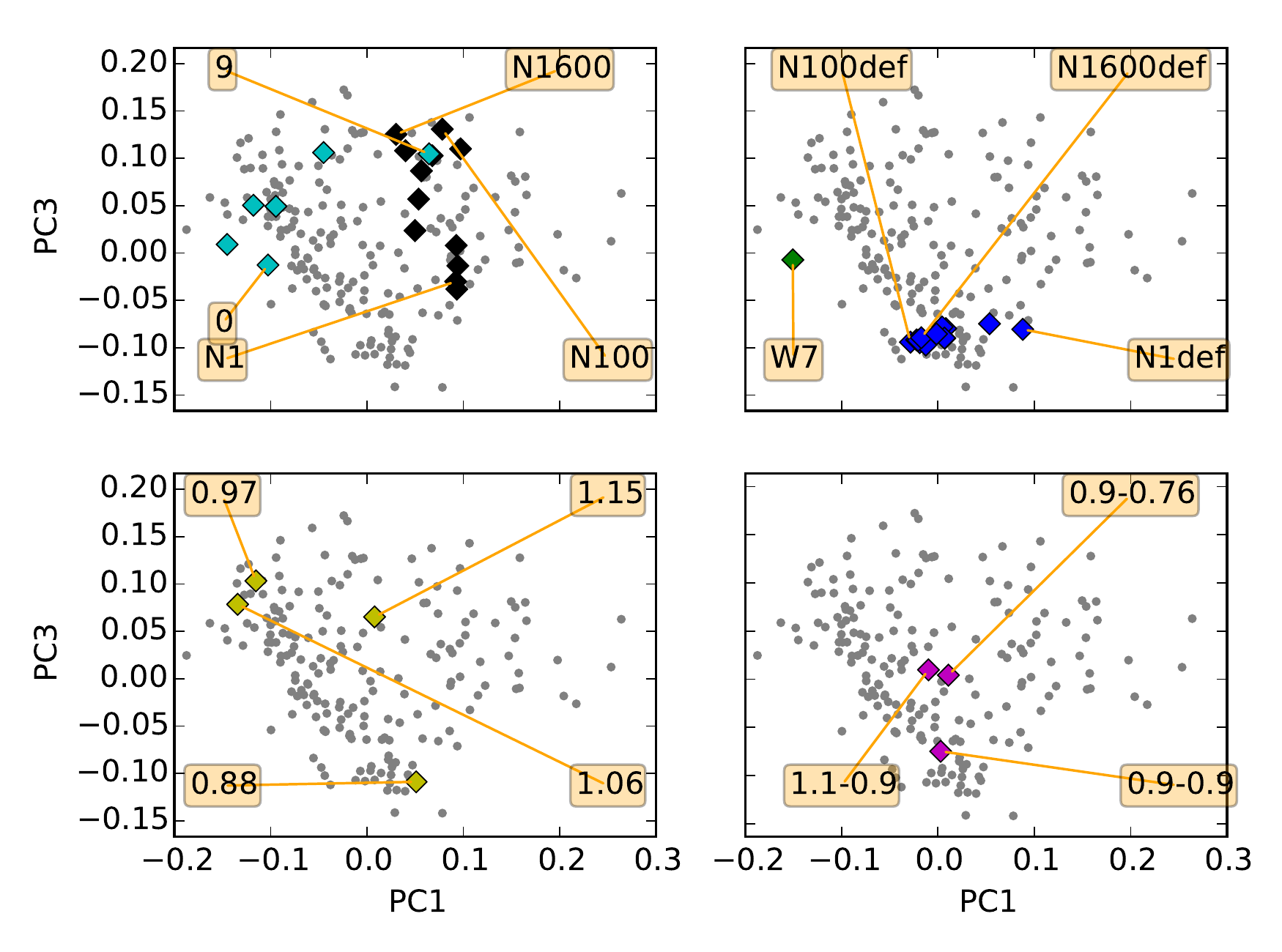}}
\caption{ The first and the third principal components of the data and
of the models over-plotted. The panels are the analogous of Fig.
\ref{fig:scatter_plot}. deflagrations (top-right), delayed detonation (top-left),
sub-Chandra (bottom-left), mergers (bottom-right). }
\label{fig:scatter_plot_2} 
\end{figure*}

\begin{table*}
{
\caption{The models used in this paper.}
\begin{tabular}{llll}
\label{tab:models}
Model names &    &   Hydro paper & Rad transport paper    \\
  &    &  &     \\ %
 Delayed detonations & & & \\
\hline
 N1,[\ldots],N1600  & & \cite{2013MNRAS.429.1156S} & \cite{2013MNRAS.436..333S} \\
\hline
\hline
 Pure deflagrations & & & \\
\hline
 N1def,[\ldots],N1600def  & & \cite{2014MNRAS.438.1762F} & \cite{2014MNRAS.438.1762F} \\
 W7  & & \cite{1984ApJ...286..644N} & \cite{2009MNRAS.398.1809K} \\
\hline
\hline
 Sub-Chandrasekhar mass detonations & & & \\
\hline
0.88, 0.97, 1.06, 1.15 & & \cite{2010ApJ...714L..52S} & \cite{2010ApJ...714L..52S} \\
\hline
\hline
Double-degenerate mergers & & & \\
\hline
0.9-0.9 & & \cite{2010Natur.463...61P} & \cite{2010Natur.463...61P} \\
1.1-0.9 & & \cite{2012ApJ...747L..10P} & \cite{2012ApJ...747L..10P} \\
0.9-0.76 & & \cite{2013ApJ...778L..18K} & \cite{2013ApJ...778L..18K} \\
\hline
\end{tabular}
}
\end{table*}

At our disposal we have a series of 3D numerical simulations for all classes of
models discussed in the previous section, to be compared with the
observations.  The models that we used and the original papers
 are summarized in Table \ref{tab:models}.
In this work we focus on the angle averaged
spectra and light curves from these models.  It will be the topic of a future
paper to use the lines of sight individually.  At first we will check the
general consistency of the spectra with observed SNe\,Ia.  To do so, we will
use only the first part of our analysis, that is the PCA space. The position
of the models in the {\color{black}first dimensions of} the PCA space gives
interesting clues about the spectral behaviour of the models.
Figs.  \ref{fig:scatter_plot} and
\ref{fig:scatter_plot_2} show where models lie in the projections on the first
three PCs. These are the most significant ones. Normalizing to the
first component, the second component has
$\sim45\%$ of the variance of the first. The third has
$\sim43\%$. The fourth component has only $\sim25\%$ of the variance of the
first component. The first five PCs cover more than 90$\%$ of the total
variance in the data. As example, the
spectrum of the W7 model at $B$ band maximum and its reconstruction
from the PCA space by means of 5 PCs is shown in Fig. \ref{fig:W7_}.
More examples are shown and discussed in Appendix B.

The model classes are characterized by the variation of a single input
parameter (the white dwarf's mass in case of sub-M$_{\text{Chan}}$
models, the number of ignition spots in case of delayed detonations and
deflagrations, and the masses of the two white dwarfs in case of
mergers). They show up in our PC space as chains of points describing
curves in the 5D space. Along these curves, the input parameter varies
continuously leading to continuous variations of the spectral
properties. Most of these models lie well within the space of observed
SNe\,Ia and cover a fair fraction of their diversity.  

\begin{figure}
\centering
\subfloat{\includegraphics[width=1.\columnwidth]{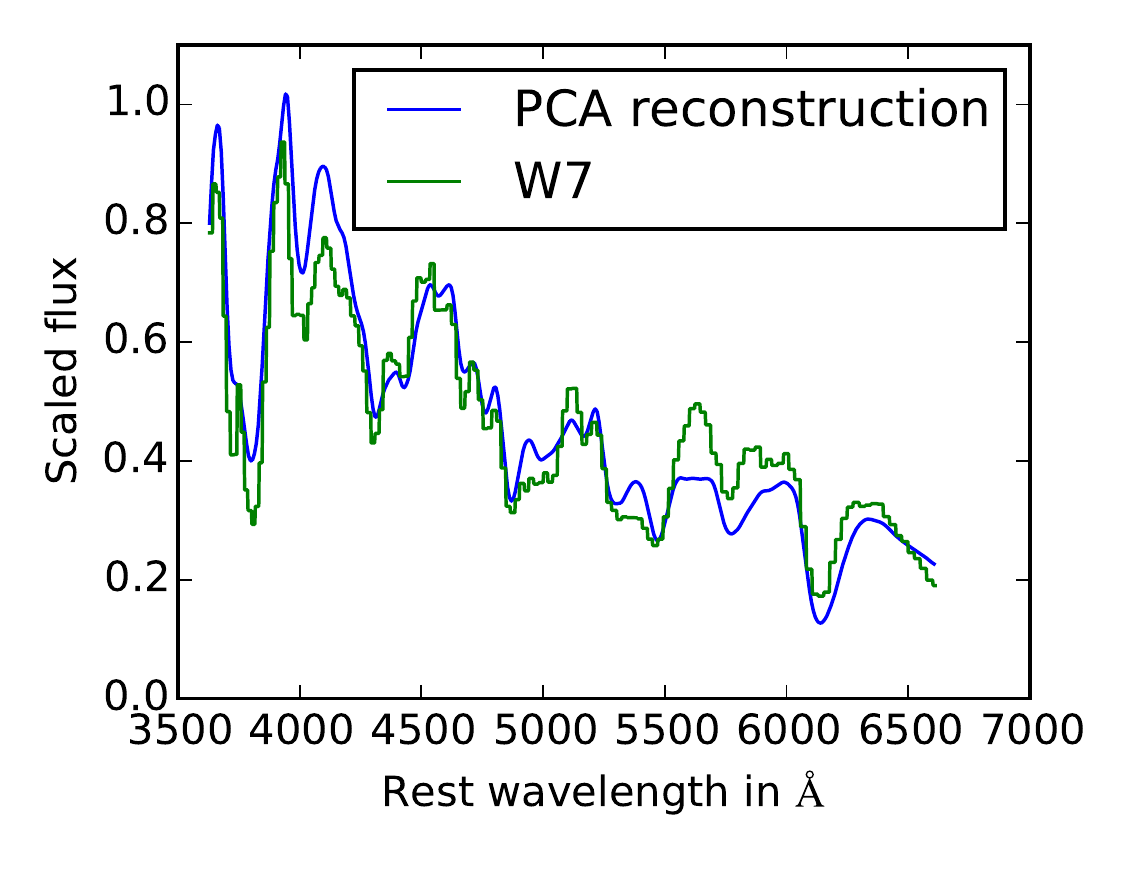}}
\caption{The figure shows the spectrum of the W7 model at $B$ maximum and the corresponding reconstruction using PCA.}
\label{fig:W7_}
\end{figure}

More specifically, we use sub-M$_{\text{Chan}}$
detonations \citep{2010ApJ...714L..52S} in the range of 0.97 to
1.15 $\Msun$. These models draw a curved line running clockwise when projected
on the first two dimensions (Fig.
\ref{fig:scatter_plot}, bottom-left panel) that connects faint 1991bg-like supernovae with
normal SNe\,Ia. The faint model with $0.88\Msun$ marks the beginning of this
line, and it is quite far away from most normal SNe\,Ia. Its next
neighbors are faint 1991bg-like and 2002cx-like supernovae. This is not
unexpected and is a confirmation of the general behaviour of the
sub-M$_{\text{Chan}}$ models \citep{2010ApJ...714L..52S}.

Our delayed detonation models \citep{2013MNRAS.429.1156S,2013MNRAS.436..333S}
lie in a completely different part of the PC space (Fig.
\ref{fig:scatter_plot}, upper-left panel).  In these delayed
detonation models the initial condition that is varied is the number of
ignition spots (N). This affects the strength of the deflagration phase.
{During the deflagration phase the rate of nuclear burning is proportional to
the surface area of the burning front. A small number of ignition spots means
that burning is less complete when the conditions for the transition to a
detonation are met.  This in turn implies less pre-expansion of the white
dwarf and a more complete burning in the detonation phase with more \Nifs\
being produced.  In the first two components the models draw a line that goes
from normal SNe\,Ia with high photospheric line velocities (N1600, $1600$ ignition
spots) down to hot and bright SNe\,Ia (N1, one ignition spot)}.  This behaviour is
also in line with what is expected from this class of
models \citep{2013MNRAS.436..333S}.

Klauser (B.A. Thesis, LMU) created a series of 1D models using a
 spherical average of N100 \citep{2013MNRAS.429.1156S} as a starting point.
 First, he constructed a completely stratified version of the
 N100 model preserving the total masses of the most important elements and the
 density profile. IGEs different from \Nifs\ are represented by stable
 Fe. Then, starting from this model he introduced progressively more mixing
 by convolving the abundances with a Gaussian window. The density profile and
 the total masses of the different elements are kept constant. This makes the
 model consistent with the total energy output.  Also, inspired
 by the results from the ``abundance tomography'' technique
\citep[e.g][]{2015MNRAS.450.2631M} the \Nifs\ is placed outside of the stable Fe.
The models add a component clearly orthogonal to the trend of
the delayed detonation models (Figs.
\ref{fig:scatter_plot} and \ref{fig:scatter_plot_2}).
From now on we call these models ``artificially mixed N100''. 
This finding shows that a mechanism which suppresses mixing in some cases can
potentially explain part of the remaining diversity of SNe\,Ia spectra
not matched by varying the number of ignition spots.

{The deflagration models from \cite{2014MNRAS.438.1762F} stay in the part of the
diagram that belongs to the faint 02cx-like SNe (Fig. \ref{fig:scatter_plot}
 top-right panel)}.
The prototypical parametrized deflagration W7 model
is closer to the bulk of the spectroscopically normal SNe\,Ia.

The two merger models with progenitors of different
mass \citep{2012ApJ...747L..10P,2013ApJ...778L..18K} are close to the
center of the distribution. The faintest merger of white dwarfs of
equal masses \citep{2010Natur.463...61P} is more separated from the
others and closer to 91bg-likes (Fig. \ref{fig:scatter_plot_2},
bottom-right panel).

In Appendix B we show how the reconstructed model spectra compare
with their original counterparts and how they compare with the nearest
observed neighbors in PCA space. We find that the agreement is quite
good in general, but becomes poor in cases where the synthetic spectra
are very different from the bulk of the data used to construct the PCA
space. However, this was expected and has to be taken into
consideration when comparing models with data. Also, one has to keep in
mind that plots like Figs. 3 and 4 are projections only and that the
``real'' distance in PCA space of a model to its observed neighbors may
be larger than it appears to be.

In the literature it is extensively discussed that in order to
have a good scenario for SNe\,Ia, the luminosity range, the luminosity-decline
rate relation and the rise time have to agree.}  But it is also important to
have consistency between spectral and photometric properties.  This will be
investigated in the following Section.

{\color{black}\section{Projecting spectral properties into photometric properties by PLS.}}

In this section we check the consistency between spectral
series and a set of well studied photometric properties with the aid of PLS
regression. The effect of dust extinction on the space of photometric
properties is marginalized by rejecting highly extinguished SNe.

\

\subsection{\Deltam}

The first relation against which we test the models is the correlation
between spectral properties and \Deltam.  In particular, we study the
empirical relation that we found in \cite{2015MNRAS.447.1247S}.  This
is analogous to the relation between the ratio of the depth of
the \SiII~5972~\AA\ and the \SiII~6355~\AA\ lines and 
\Deltam\ \citep{1995ApJ...455L.147N}, but studied with the
systematic approach of the PLS. We stress that the relation
explored in this section is not equivalent to the Phillips-relation and it is an
additional constraint on the models. It is a relation between global
spectral properties (predicted \Deltam) and light-curve decline
(observed \Deltam).

\begin{figure*}
\centering
\subfloat{\includegraphics[width=2.\columnwidth]{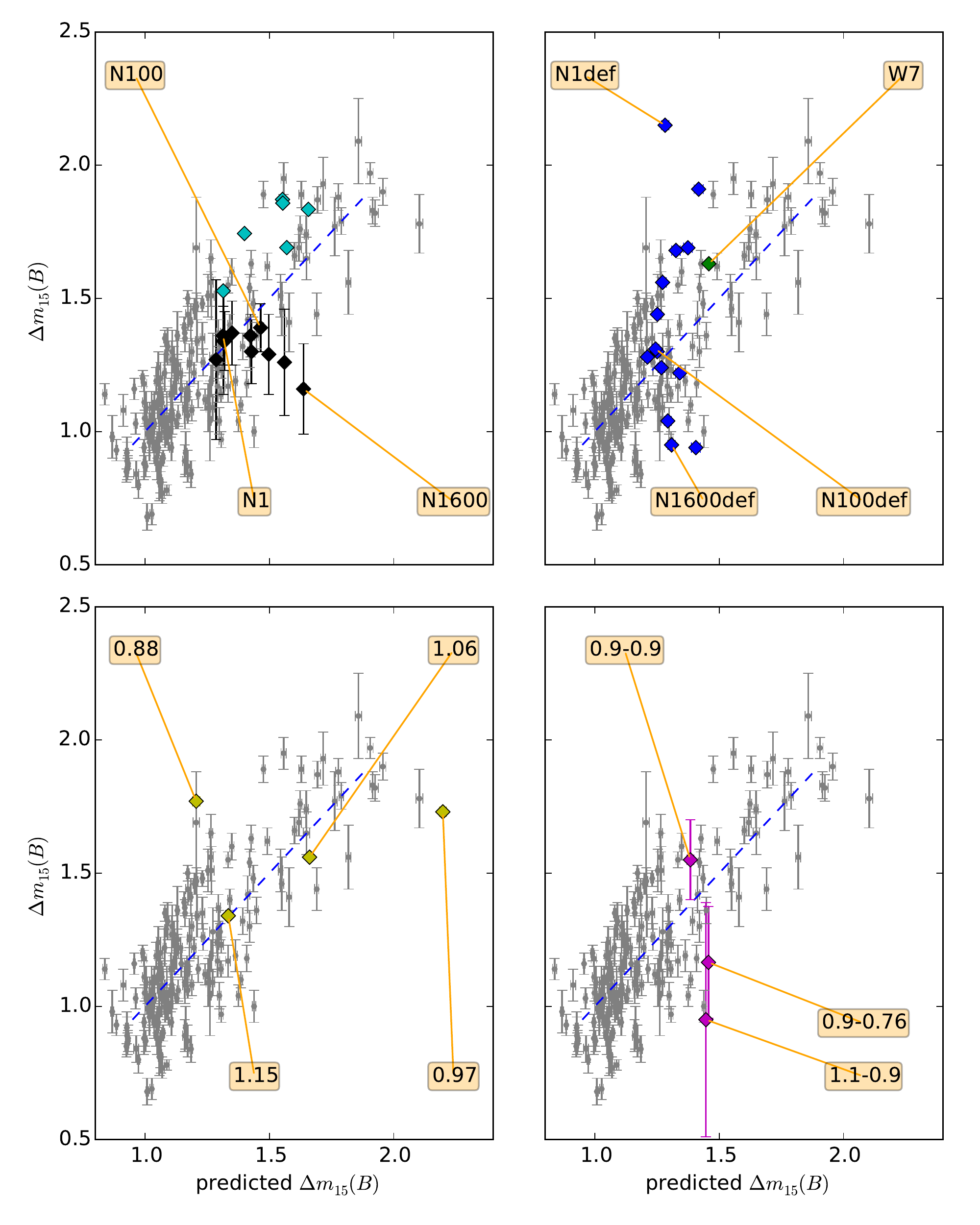}}
\caption{The grey dots show the relation of \Deltam\  vs. the 
predictions obtained by using PLS on the PCA space of
spectra for observed SNe\,Ia. The models are deflagrations (top-right), delayed detonation
(black) and modified-mixing models (cyan) (top-left), sub-Chandra
(bottom-left), mergers (bottom-right).  Errors on the predictions come
from k-folding \citep{2015MNRAS.447.1247S} and are smaller than the size of the diamonds.
 The variance on the observed
\Deltam\ of mergers and delayed-detonation models comes from the
variability due to line-of-sight effects.}
\label{fig:DM15_models}
\end{figure*}

In Fig. \ref{fig:DM15_models} we show the correlation between  
\Deltam\ and the direction in the PCA space that correlates with it. It is
found by PLS and by using public data only. This direction is close to
the direction that predicts the equivalent width of \SiII~5972~\AA\
\citep[][Figure 12]{2015MNRAS.447.1247S}.
The projections of models representative for the different scenarios
are over-plotted.

In the interval between
0.97 $\Msun$ and  1.15 $\Msun$ the sub-Chandrasekhar mass detonation
models reproduce the fast decline part of the relation remarkably well (yellow diamonds
in Fig. \ref{fig:DM15_models}). This is not too
surprising since it was shown in \cite{2010ApJ...714L..52S} that they
follow the Phillips-relation. However, they do not reach the slow decline part
of the relation. 

In principle, the parameter space of the merger model is large and with
the three models available to date we can just begin to explore it.  The
brightest merger (1.1 and 0.9 $\Msun$) is clearly below the empirical
relation, particularly for some line of sights
(indicated by the error bars). This means
that, for the given spectral properties of the model, its light curve
evolution is too slow \citep{2012ApJ...747L..10P,2012ApJ...750L..19R}. In
turn, this implies that the opacity of the model
is too large which slows down the evolution of the light curve.  A
likely explanation is that the total mass is too large to reproduce the
bulk of normal SNe\,Ia. This interpretation is confirmed by the
qualitatively similar merger model (0.9 and 0.76 $\Msun$) which matches
the relationship much better, however, this model cannot be considered a
good match for normal SNe\,Ia from a luminosity or spectroscopic point of view 
(see section \ref{subsec:B_mag}). The equal-mass merger (0.9 and
0.9 $\Msun$) lies also on the relation. 

W7, the classical 1D deflagration model
from \cite{1984ApJ...286..644N}, that is a good match for
spectroscopically normal SNe\,Ia at early epochs, lies
well on the relation. However, the more realistic deflagration
models do not match spectroscopically with the bulk of SNe\,Ia (see
section \ref{sec:PCAspace}).  This is why the PLS regression, that
works well in predicting the \Deltam\ for the bulk of SNe\,Ia, returns
a value close to the average \Deltam\ for these objects. A richer
training sample including more faint objects is necessary to study
such a scenario with PLS.

The 3D delayed detonation
models from \cite{2013MNRAS.429.1156S} and \cite{2013MNRAS.436..333S} 
cluster in a single area of the empirical relation, but do not show 
its observed trend. Since in the models the number of 
ignition spots was used to parametrize the strength of the initial
deflagration this means that at least one more parameter is needed
to reproduce the relation within this class of models. Moreover,
models with low or moderate deflagration strength (N lower than about 300)
are within the observed range. 

Finally, the models plotted as cyan diamonds in Fig. \ref{fig:DM15_models}
are the artificially mixed N100 models
Klauser (B.A. Thesis, LMU). In contrast to the other
delayed-detonation models they do follow the empirical relation and
models with a low degree of mixing have larger predicted and
observed \Deltam.  This may indicate that a mechanism that allows for
more stratified ejecta may be needed in order to reproduce the
observed correlation between \Deltam\ and spectral properties with this
class of models.  Rayleigh-Taylor instabilities in the 
equatorial plane are suppressed by rotation.
Hence, rotation is a possible mechanism that may help
in suppressing mixing. \cite{2010A&A...509A..74P,2010A&A...509A..75P} 
studied explosions of differentially rotating white dwarfs and concluded 
that they can not be a good explanation for normal SNe\,Ia. However, 
systematic studies of rigidly rotating white dwarfs are still missing.

\subsection{$B-V$ colors}

\begin{figure*}
\centering
\subfloat{\includegraphics[width=2.\columnwidth]{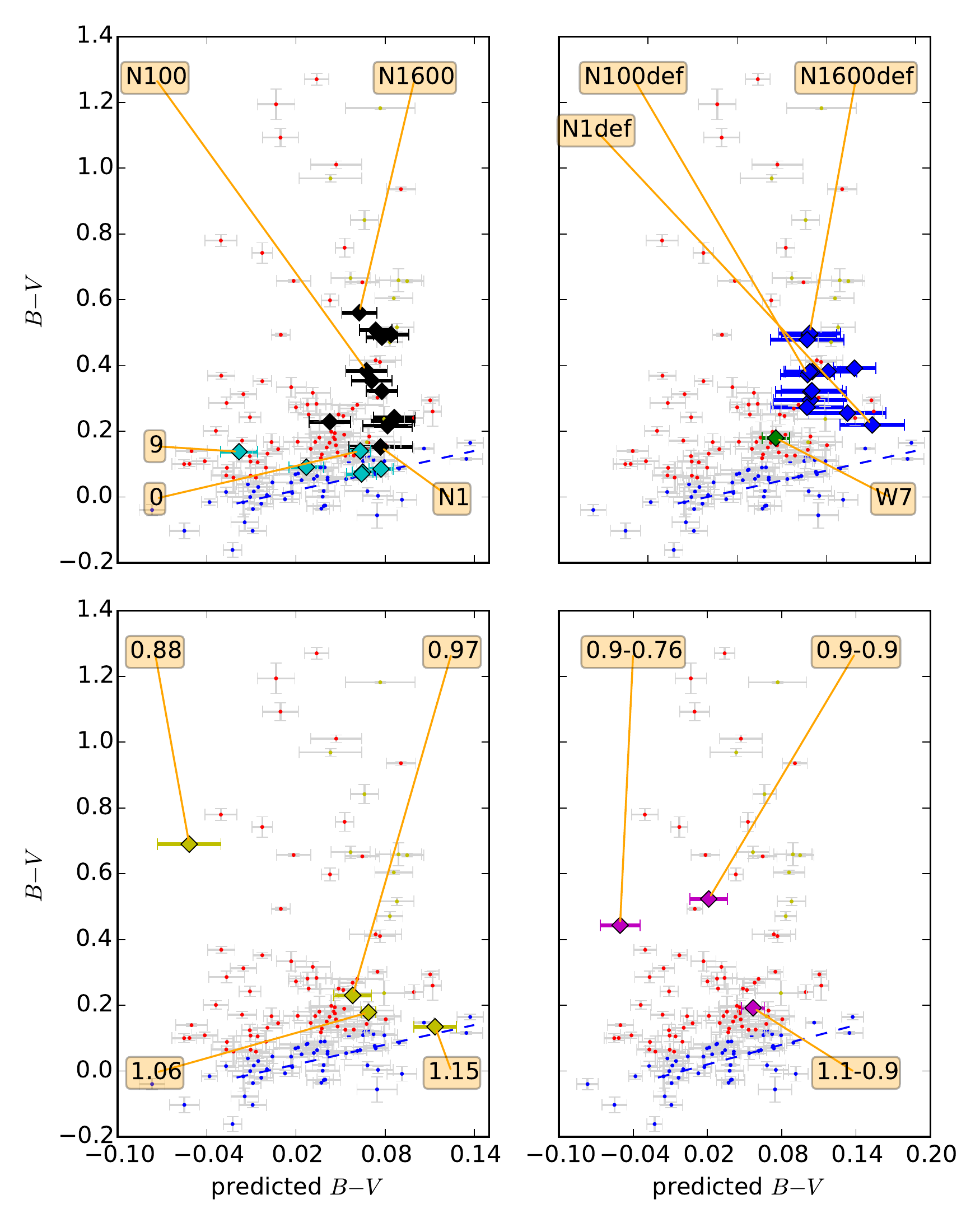}}
\caption{The $B-V$ color at $B$-max (without reddening corrections) and the
$B-V$ color predicted from spectral properties using PCA and PLS.  The SNe are
represented by dots. SNe with  small reddening are in blue, the SNe with
significant reddening are in red.  Errors on the predicted
colors come from k-folding \citep{2015MNRAS.447.1247S} on the PLS regression
analysis. Only the blue points are representative of the intrinsic SN
properties,
the relation is the blue dashed line. SNe in yellow belong to a part
of the space with not enough photometric data. For them the predicted color is
not reliable.  The models are represented by diamonds. The models are W7 and 3D
deflagrations
(top-right), delayed detonation  and modified-mixing models  (top-left),
sub-Chandra (bottom-left), mergers (bottom-right).} 
\label{fig:BmV_models} 
\end{figure*}

Next we study the consistency of the $B-V$ color of the models at
$B$-band maximum with the observations. The relation between color and
spectral properties is similar to the relation between color and
velocity of \SiII~6355~\AA\ \citep{2011ApJ...729...55F}. Once again,
our approach allows for a systematic study of this property and it
allows us to use all the information present in the spectra at
different epochs, and not only the behaviour of the spectra at $B$-band
maximum.

An important remark here is that in principle our analysis is valid
for all SNe\,Ia. This includes spectroscopically normal SNe\,Ia,
1991T-likes, those with broad lines \citep{2006PASP..118..560B}, and
others. But it is not possible to study their color without enough
photometric data. In particular the 1991bg-like and the
2002cx-like events present in the sample, although they cluster nicely in
different parts of the PCA space (see Section~\ref{sec:PCAspace}), do not have well
enough measured photometry to study their colors with the PLS analysis.

To quantify for which SNe we can reliably predict the colors we
check if a given SN has neighbours which are selected to have negligible reddening by
the PLS algorithm.  In Fig. \ref{fig:BmV_models} the SNe are colored in blue if
they are selected by the PLS algorithm and have negligible reddening.  They are
colored in red if they are not selected as unreddened by the PLS algorithm, but
have close neighbours that are unreddened. This means that the color excess is
likely due to reddening.  The SNe that do not have a neighbour selected as
unreddened are colored in yellow. They are 1991bg-like and 2002cx-like. For
those we cannot predict their color or luminosity using PLS.

 With this in mind, the empirical relation between intrinsic color and
spectral properties holds for the bulk of SNe\,Ia, but
not for most peculiar SNe.  Therefore the relation has to hold for
models proposed to describe the bulk of SNe\,Ia but not for rare
objects.  {\color{black} Of course, by construction, the models
are not reddened. Only models that fit in the relation defined by
the SNe colored in blue can explain the bulk of SNe\,Ia.  Ideally, a
perfect scenario for normal SNe\,Ia has to lie in the strip populated
by blue dots in Figs \ref{fig:BmV_models} and
\ref{fig:Bmag_models}.}
Some of the  models that are off this relation can be a good explanation for
some of the peculiar SNe colored in yellow. For these SNe the PLS predicted 
quantity is just a linear extrapolation of what is valid for normal SNe.
Until a significantly larger sample is available, 
these peculiar SNe need to be studied on a
case-by-case basis.

Many of the models suffer from being too red compared to the
observations (Fig. \ref{fig:BmV_models}). This systematic issue is
well known \citep[e.g][]{2013MNRAS.436..333S} and it may be due
to approximations in the
radiation transport code or shortcomings in the hydrodynamic modelling.  Here,
we focus on the trend between
intrinsic color and spectral properties that holds for the majority of
observed SNe\,Ia, but does not seem to be clearly reproduced by any
of the investigated explosion scenarios for which more than one
realization exists.

The intrinsic color of SNe\,Ia correlates with the velocity of the main
\SiII\ line \citep{2011ApJ...729...55F}. SNe\,Ia with a higher velocity have 
redder color.
Consequently, the delayed detonation models, well known for having high photospheric
velocities \citep{2013MNRAS.436..333S}, place themselves in
the right side of the diagram in Fig. \ref{fig:BmV_models} (top-left panel),
 where the intrinsic color is larger
than $\sim0.1$. However, most of these models are still too red. For
example, N100, {\color{black}a delayed detonation model proposed as explanation 
for normal SNe\,Ia}, is too red
by $\sim0.3$ mag. It is interesting to note that models with the same
composition, but with an enforced stratification, reproduce the trend of the
observed relation fairly well. The crucial difference in obtaining the right color
can be due to the stratification of the inner parts of the
ejecta. These modified models have stable iron at the center
and \Nifs\ around it. This changes the way the light is reprocessed
and makes the color less red. We also note
that the 2D delayed detonation models of \cite{2009Natur.460..869K}
are systematically bluer than our models. However, it is unclear to
which extent this is due to the reduced dimensionality of the models
differences in the radiative transfer or nucleosynthesis postprocessing.

Both the sub-M$_{\text{Chan}}$ models and the delayed detonation models
 show a trend that is opposite to the one
observed for normal SNe\,Ia. Brighter models have IME at higher
velocities. This trend seems to be a robust
characteristic of these scenarios.
For the sub-M$_{\text{Chan}}$ models more massive progenitors will
naturally have higher kinetic energies and ejecta opaque up to higher
velocities. Also, the IME layer moves at higher velocities because
more IGEs are produced near the center. At the same time, more massive and
brighter models, within the range of luminosities of typical SNe\,Ia,
are going to be hotter and thus bluer. Among normal SNe\,Ia, those with lower
photospheric velocities can have both, greater or lower luminosity.
For the delayed detonations more complete burning releases more
energy. If the binding energy is the same, the kinetic energy will be larger.

As previously discussed, there are not enough data available to study
the color of 1991bg-like SNe with PLS regression.  However, peculiar
1991bg-like supernovae are known to have low photospheric velocities
and to be intrinsically redder than normal ones \citep{1996MNRAS.283....1T}. For them,
the sub-M$_{\text{Chan}}$ models with a low mass may be a viable scenario.

The color of the brightest merger model considered is quite right,
just a bit too red (Fig. \ref{fig:BmV_models}, bottom-right panel),
and this could be due to issues in the radiation transport and not in
the scenario itself.  We need merger models with initial conditions
similar to the brightest merger to check if this scenario reproduces
the trend observed for normal SNe\,Ia.  The merger models with lower
masses are too far away from the properties of normal SNe\,Ia to be
used to extrapolate such a relation.

\subsection{$B$ magnitudes}

\label{subsec:B_mag}

\begin{figure*}
\centering
\subfloat{\includegraphics[width=2.\columnwidth]{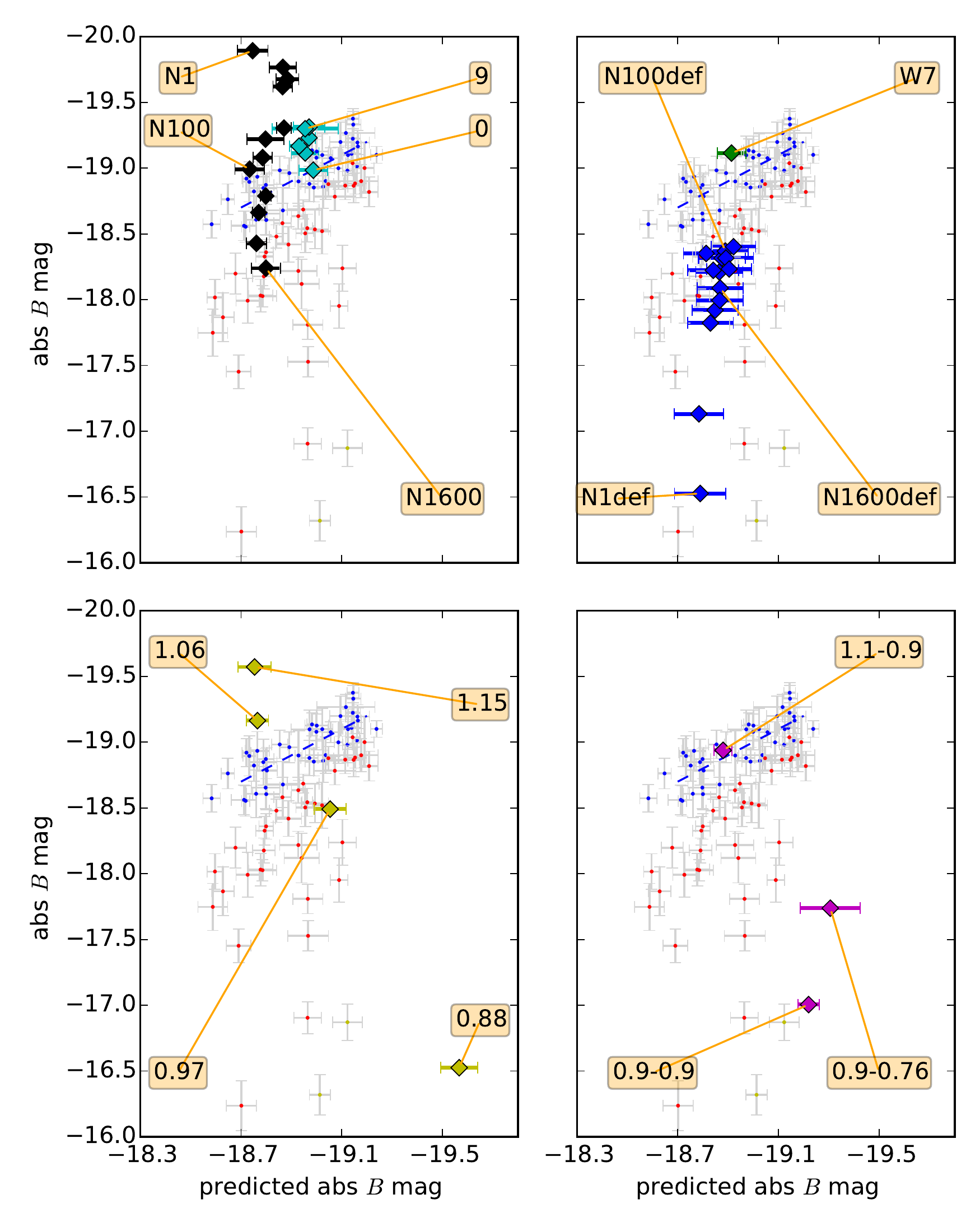}}
\caption{The absolute $B$-band maximum without reddening corrections and
the predicted $B$max using PCA and PLS. The models are W7 (top-right, green),
3D deflagrations (top-right, blue),
delayed detonation (black) and modified-mixing models (cyan)
(top-left), sub-Chandra (bottom-left), mergers (bottom-right). Errors
on the predictors come from k-folding on the PLS regression analysis
\citep{2015MNRAS.447.1247S}. (See Fig. \ref{fig:BmV_models}).
The models that are far away from the blue strip can
not explain normal SNe\,Ia. However, some can be good candidates for some
peculiar SNe\,Ia (e.g. SNe marked in yellow).
}
\label{fig:Bmag_models} 
\end{figure*}

In order to study the relation between spectral properties and absolute
magnitudes we need to have an estimate of the absolute magnitude,
independent of the assumption of a width-luminosity relation of 
the light curves.  Assuming a
Hubble constant of $70$\kmsMpc, we use the redshift as a measure of
distance. The error of the absolute magnitude is then the error on the
observed magnitude with an error due to the peculiar motions of the
galaxies added in quadrature.  We do not attempt to perform any
reddening correction. Similarly to the case of the colors, we can not
study the faintest SNe\,Ia with statistical methods since we do not have
enough 1991bg-like and 2002cx-like supernovae in the smooth Hubble flow.

In Fig. \ref{fig:Bmag_models} sub-M$_{\text{Chan}}$ models with initial masses
between $0.97$ and $1.15\Msun$ bridge the correct range of luminosities of the
bulk of SNe\,Ia. However, they are orthogonal to the empirical relation between
luminosity and spectral properties. Among the brightest representatives of the
SN\,Ia population are 1991T-like SNe. They are fairly common and are 
characterized by spectra with high temperature and somewhat
lower than normal photospheric velocities. This is in contrast to what the
sub-M$_{\text{Chan}}$ models predict. SNe with ``normal'' luminosities show a
diverse range of photospheric velocities that is correlated with color, but
not with luminosity. To explain the variability shown in the spectra a
parameter other than the mass at explosion is needed in the
sub-M$_{\text{Chan}}$ scenario. This parameter needs to decrease the \Nifs\
mass of the ejecta to slow down the time evolution without changing the total
mass.

Fig. \ref{fig:Bmag_models} shows that delayed detonation models can also
explain the appropriate range of luminosities easily but, like the
sub-M$_{\text{Chan}}$ models they are mostly orthogonal to the observed
relation. The behaviour of models with limited mixing suggests that a
mechanism to suppress mixing in the ejecta can possibly bring the delayed
detonation models into closer agreement with the observations. Models with
high stratification in the ejecta go towards the right side of the diagram. In
order to explain the bright SNe\,Ia a high stratification of the ejecta seems to
be necessary. Significant rotation of the progenitor may be a possibility
to suppress mixing and to produce more stratified ejecta.

{The W7 model is placed close to the relation.  The modern 3D deflagration
models, on the other hand, are significantly fainter than the bulk of SN\,Ia,
and can not be used to explain them.  Their luminosity, however, is compatible
with some of the faint classes such as 02cx-like SNe.}

The brightest merger model lies nicely on the relation observed for
the bulk of SNe\,Ia. The lower mass mergers are too faint for such a
relation, and they cannot be an explanation for it. However, they can
explain fainter and rarer objects. To assess if the merger scenario is
a viable explanation for the majority of SN\,Ia, one has to explore
the parameter space close to the bright $1.1$ and $0.9\Msun$ merger
model to see if the relation holds.  The faintest merger models can
not be a good explanation for the normal SNe\,Ia. Their $B$-band
luminosity, as predicted from their spectral properties, would be far too
high (see Fig. \ref{fig:Bmag_models}). The predicted $B$~mag is just
a projection of the PCA space, and it is a good prediction for the
$B$~mag for the bulk of SNe\,Ia only.  The fact that these models are
off such a relation does not mean that they do not provide an
explanation for some peculiar objects
\citep[e.g.][]{2013ApJ...778L..18K}.

\section{Conclusions}
\label{sec:conclusions}

 \cite{2015MNRAS.447.1247S} studied the empirical relations between
the small wavelength spectroscopic properties of SNe\,Ia and
photometry.  In this work we test a number of models for these
relations.  Our tool allows a systematic comparison between synthetic
and real SN~Ia spectra.  We use PCA on time series of observed SN\,Ia
spectra to reduce the dimensionality of the data.  Projecting the
synthetic spectral series of a particular model in the PC space
revealed its counterparts among the real supernovae by an analysis of
its neighbours. Moreover, the relations discovered by PLS are strong
tests that the models have to pass.  Given the challenge of performing
a coherent statistical comparison between synthetic and real spectra,
our method is particularly efficient in characterizing large sets of
models built from different explosion scenarios. It is able to provide
important insights regarding the global properties of each explosion
mechanism in order to favour or disfavour them. Such a global analysis
is also expected to be more robust against systematics in the models
(for example due to approximations in the radiation transport) than
comparing them individually on a case-by-case basis with real SNe.

Much of the known behaviour of the models is recovered in our work.
For example, the pure deflagration models can be an explanation for
faint SNe~Ia
\citep{2012ApJ...761L..23J,2013MNRAS.429.2287K,2015MNRAS.450.3045K,2015A&A...573A...2S}.
Similarly, also the merger models with equal initial
masses (0.9-0.9 $\Msun$) are a good candidate for faint SNe~Ia
 \citep{2010Natur.463...61P}.  The
sub-Chandrasekhar mass detonations and the delayed detonations are the
best candidates for normal
SN~Ia \citep{2010ApJ...714L..52S,2011MNRAS.417..408R}.
However our technique shows that
they both have shortcomings that need to be addressed.  The number of
merger models studied is very limited, and only one is a good
candidate for normal SNe\,Ia.  So it is hard to put strong constraints
on this scenario.

  In addition to what was known before, our tool offers very stringent
tests for those models that are candidates for the bulk of SNe~Ia,
thanks to the abundant observational data.  We find that the relations
between spectroscopic and photometric properties predicted by the
sub-M$_{Chan}$ models do not follow the relations that are valid for
the bulk of SN~Ia. On the other hand, sub-M$_{Chan}$ models with
masses lower than 0.97 $\Msun$ are very good candidates for the faint
SN~Ia, where such relations are reproduced.  Some of the shortcomings
of the delayed detonation models as an explanation for the bulk of
SNe~Ia may be cured by a mechanism which reduces mixing in the
brightest models.  We propose rotation as a possibility to achieve
this result.  A merger model with masses (1.1-0.9 $\Msun$), proposed
as an explanation for the bulk of SNe~Ia, evolves somewhat too slowly
in relation to its spectral properties. This suggests an ejecta mass
that is somewhat too high. On the other hand, it is positioned quite
well in the diagrams that show luminosity or color. An investigation
of more merger models is necessary to reach definitive conclusions for
this scenario.

Once a large enough library of synthetic spectra becomes available, an
alternative PC space may be constructed directly from the
models. Projecting the observed SNe to such a model PC space, will
provide a critical cross check between the real and synthetic metric
spaces. In addition, our approach is well suited to
systematically assess viewing-angle effects in multi-dimensional
explosion models, which is very difficult on an individual basis. Both
 of these aspects will be addressed in future work.


\section*{Acknowledgments}

The authors gratefully acknowledge the Gauss Centre for Supercomputing (GCS)
for providing computing time through the John von Neumann Institute for
Computing (NIC) on the GCS share of the supercomputer JUQUEEN 
\citep{stephan2015juqueen} at J\"{u}lich
Supercomputing Centre (JSC). GCS is the alliance of the three national
supercomputing centres HLRS (Universit\"{a}t Stuttgart), JSC (Forschungszentrum
J\"{u}lich), and LRZ (Bayerische Akademie der Wissenschaften), funded by the German
Federal Ministry of Education and Research (BMBF) and the German State
Ministries for Research of Baden-W\"{u}rttemberg (MWK), Bayern (StMWFK) and
Nordrhein-Westfalen (MIWF).
The work of FKR is supported by the ARCHES prize of the German Ministry of
Education and Research (BMBF) and by the DAAD/Go8 German-Australian exchange
programme for travel support.
The work of FKR and RP is supported by the Klaus Tschira Foundation.
EEOI is partially supported by the Brazilian agency CAPES (grant number 9229-13-2).
IRS was supported by the Australian Research Council Laureate Grant FL0992131.
RP acknowledges support by the European Research Council under ERC-StG grant EXAGAL-308037.
WH acknowledge support by project TRR 33  ’The  Dark  Universe’
of  the  German  Research  Foundation  (DFG) and the Excellence Cluster
``Origin and Structure of the Universe'' at the Technische Universit\"at
M\"unchen.

\label{lastpage}

\appendix

\section{PCA reconstruction of observed supernovae}
\label{sec:appendix_a}

In this Appendix we show spectra of several supernovae from our sample
together with their PCA reconstruction. To demonstrate the power (and
potential weaknesses) of the method, we have chosen examples from the
various groups in Figs. \ref{fig:scatter_plot_sne} and
\ref{fig:scatter_plot_sne_2}, i.e., ``normal'', ``HV SiII'',
``91T-like'', ``91bg-like'', and ``peculiar'', as well as some with good or
poor time coverage, and some with good or noisy data. In all figures
that follow the logarithm of flux over wavelength is plotted for the
wavelength range used in our EMPCA. Each spectrum is labeled with the
time measured from B-band maximum, and time is progresing from bottom
to top. Observed spectra are in green and reconstructed spectra are in
blue. The reconstruction is done with the first 5 principle
components. Note that in our analysis all epochs are binned. The bins
are 2.5 days wide. The curves in the figures are labled with the
midpoints of each bin.

\begin{figure*}
\centering
\subfloat{\includegraphics[width=1.2\columnwidth]{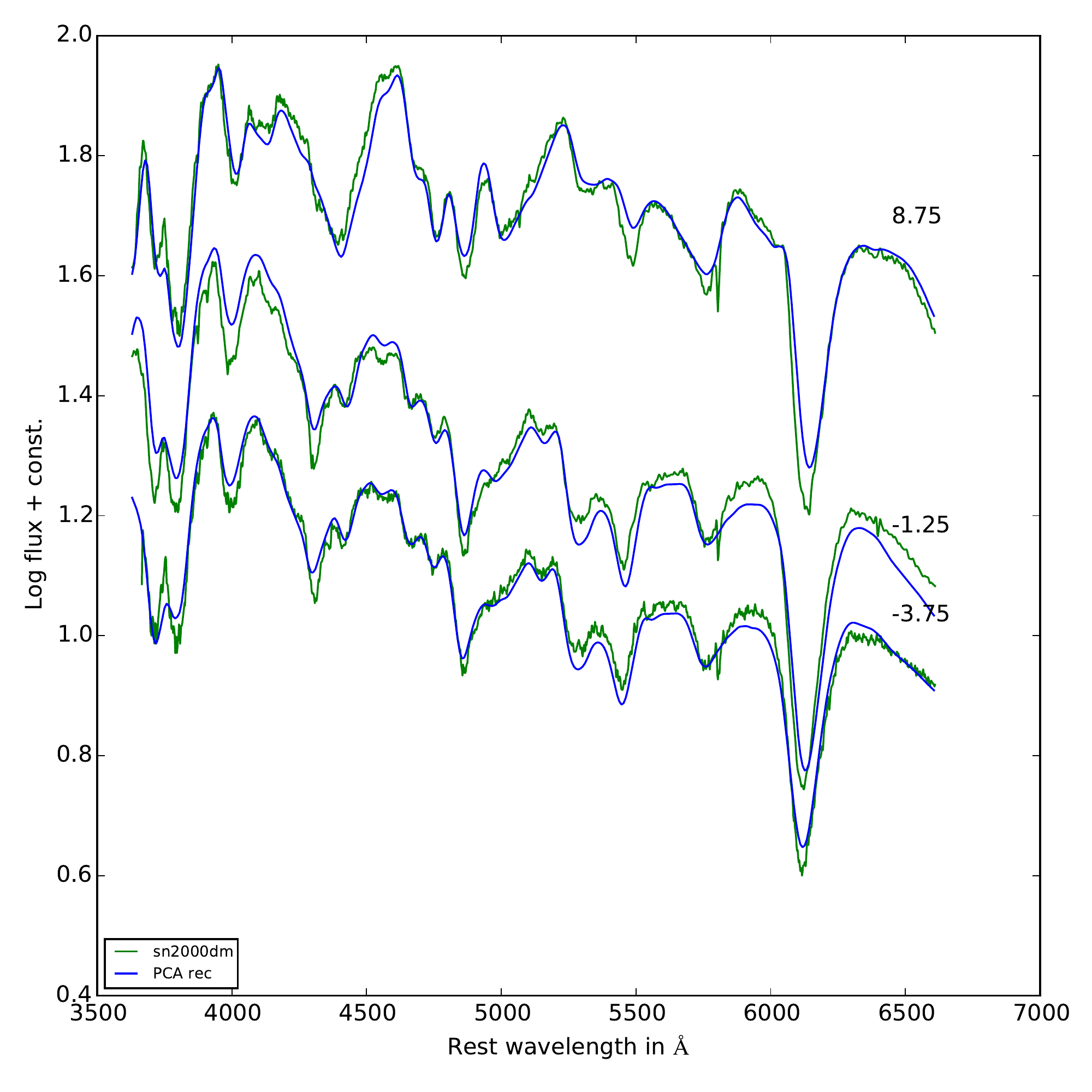}}
\caption{Observed spectra of SN 2000dm (in green) with the
  reconstructed spectra (in blue) overplotted. The data are from
  \protect\cite{2012AJ....143..126B} and
  \protect\cite{2012MNRAS.425.1789S}. }
\label{fig:spectra_sn2000dm}
\end{figure*}

We start by discussing a few spectroscopically normal SNe\,Ia. Fig.
\ref{fig:spectra_sn2000dm} shows SN 2000dm in UGC 11198 discovered by
the KAIT LOSS team, for which we have three spectra
only. Nevertheless, the reconstructed spectra agree reasonably well
with their observed counterparts at all epochs. In more general terms,
some noise or some missing epochs are not a problem if the epoch
coverage spans the range of the analysis.

\begin{figure*}
\centering
\subfloat{\includegraphics[width=1.2\columnwidth]{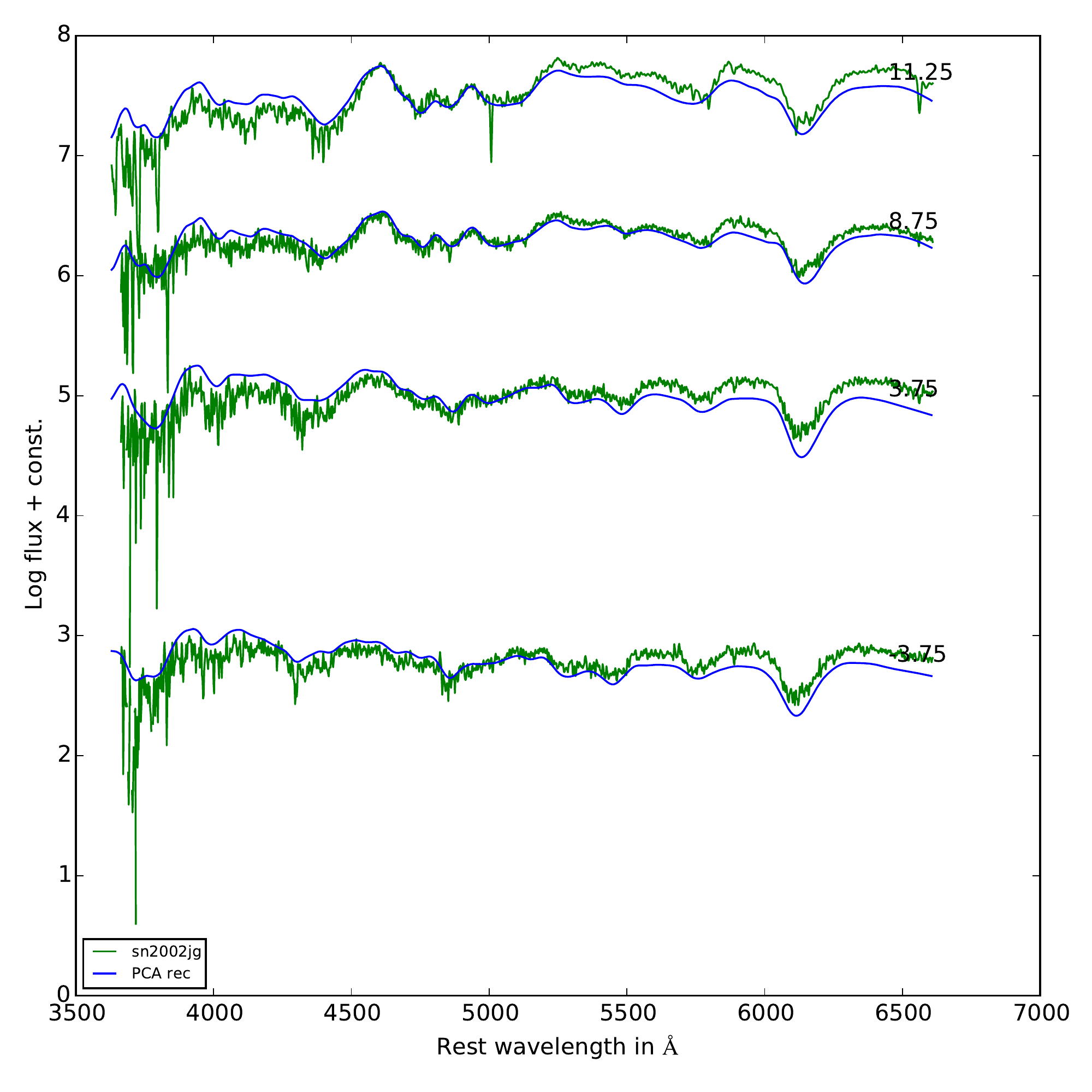}}
\caption{Observed spectra of SN 2002jg (in green) with the
  reconstructed spectra (in blue) overplotted. The data are from
  \protect\cite{2012AJ....143..126B} and
  \protect\cite{2012MNRAS.425.1789S}. }
\label{fig:spectra_sn2002jg}
\end{figure*}

\begin{figure*}
\centering
\subfloat{\includegraphics[width=1.2\columnwidth]{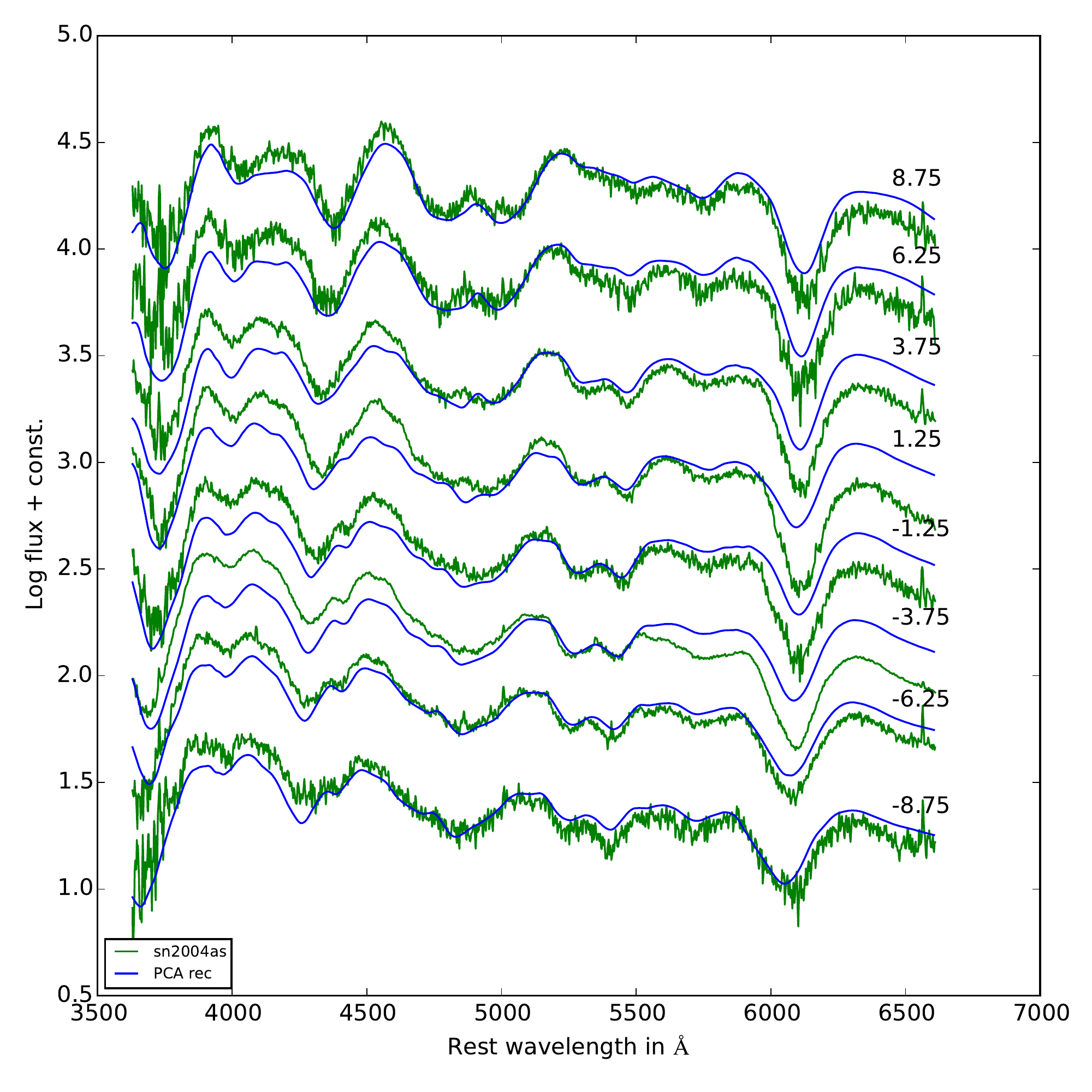}}
\caption{Observed spectra of SN 2004as (in green) with the
  reconstructed spectra (in blue) overplotted. The data are from
  \protect\cite{2012AJ....143..126B} and
  \protect\cite{2012MNRAS.425.1789S}. }
\label{fig:spectra_sn2004as}
\end{figure*}

The second example, SN 2002jg in NGC 7253, also discovered by KAIT, is
another normal SN Ia, but with more noisy
data. Fig. \ref{fig:spectra_sn2002jg} shows the four observed spectra
and their reconstruction. The fit is not affected by the data
quality. However, it can be seen that the color of the reconstructed
spectra is slightly too blue at all epochs, although all spectral
features are well reproduced. Such a slight missmatch of colors is
seen in several of our reconstructed spectra and is caused by the fact
that our analysis is based on the derivatives of the logarithm of the
spectra making the predicted colors somewhat uncertain. In addition,
reddening does not significantly affect the PCA space. This means that
the color might be poorly reproduced when there is extinction. Since,
however, the comparsion of models with data is done by means of
derivative spectroscopy this shortcoming is not a major problem. The
third example, SN 2004as, shown in Fig. \ref{fig:spectra_sn2004as}, is
an extreme case in this respect. Here, the reconstructed spectra are
too red. However, as before, individual line shapes agree quite well
at all epochs. In total, we had 16 spectra in this case.

\begin{figure*}
\centering
\subfloat{\includegraphics[width=1.2\columnwidth]{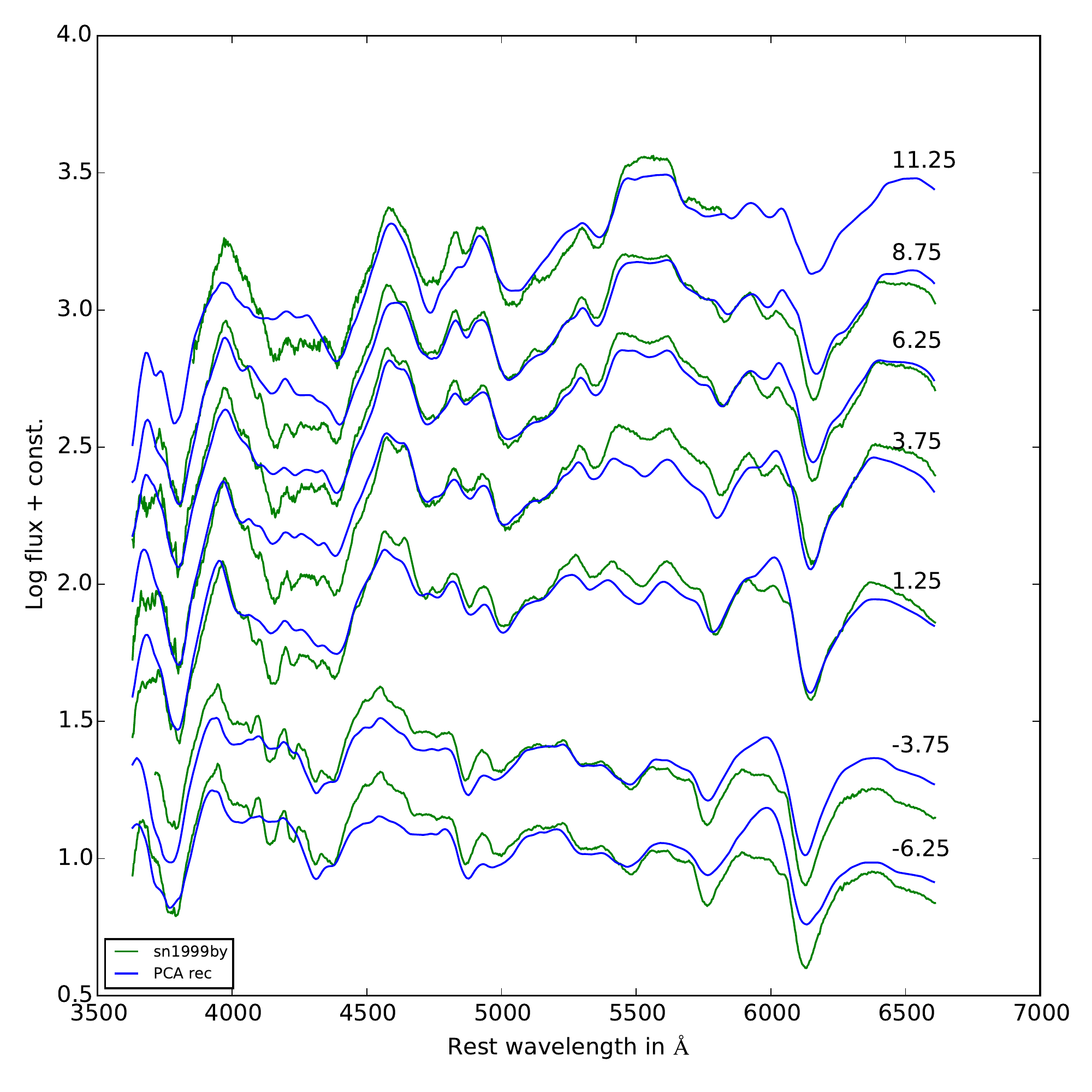}}
\caption{Observed spectra of SN 1999by (in green) with the
  reconstructed spectra (in blue) overplotted. The data are
  mostly from  \protect\cite{2012AJ....143..126B}.}
\label{fig:spectra_sn1999by}
\end{figure*}

\begin{figure*}
\centering
\subfloat{\includegraphics[width=1.2\columnwidth]{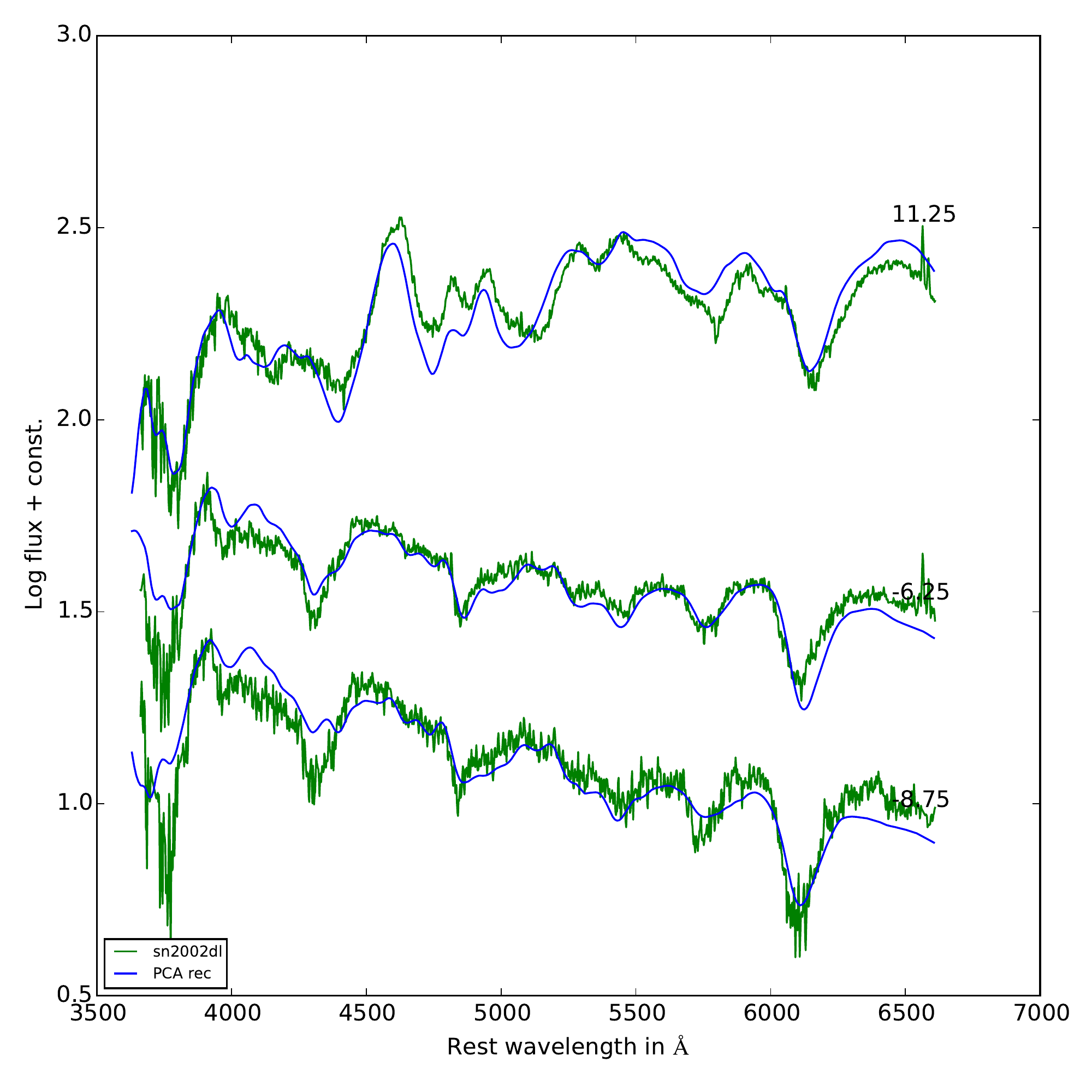}}
\caption{Observed spectra of SN 2002dl (in green) with the
  reconstructed spectra (in blue) overplotted. The data are from
  \protect\cite{2012AJ....143..126B}. }
\label{fig:spectra_sn2002dl}
\end{figure*}

Next, we discuss sub-luminous 91bg-like supernovae. Since there are
fewer members of this group in our training set, we expect that
typically the reconstructed spectra may not fit their observed
counterparts as well as they did for the normal ones, and the results
confirm this expectation. Fig. \ref{fig:spectra_sn1999by} shows
spectra of the well-observed 91bg-like supernova SN 1999by for which
we have a total of 22 spectra. Although the main features are
reproduced, the overall fit is not too good. A significantly better
reconstruction is found for SN 2002dl
(Fig. \ref{fig:spectra_sn2002dl}) for which (again with slightly
incorrect colors) the fit is very good although the data are rather
noisy and we had 4 spectra only. SN 2002dl is a less ``extreme''
91bg-like and this is why it is easier to reproduce.  All in all,
however, most spectral features of 91bg-like SNe Ia are reasonably
well reproduced and some of the shortcomings are not unexpected.

\begin{figure*}
\centering
\subfloat{\includegraphics[width=1.2\columnwidth]{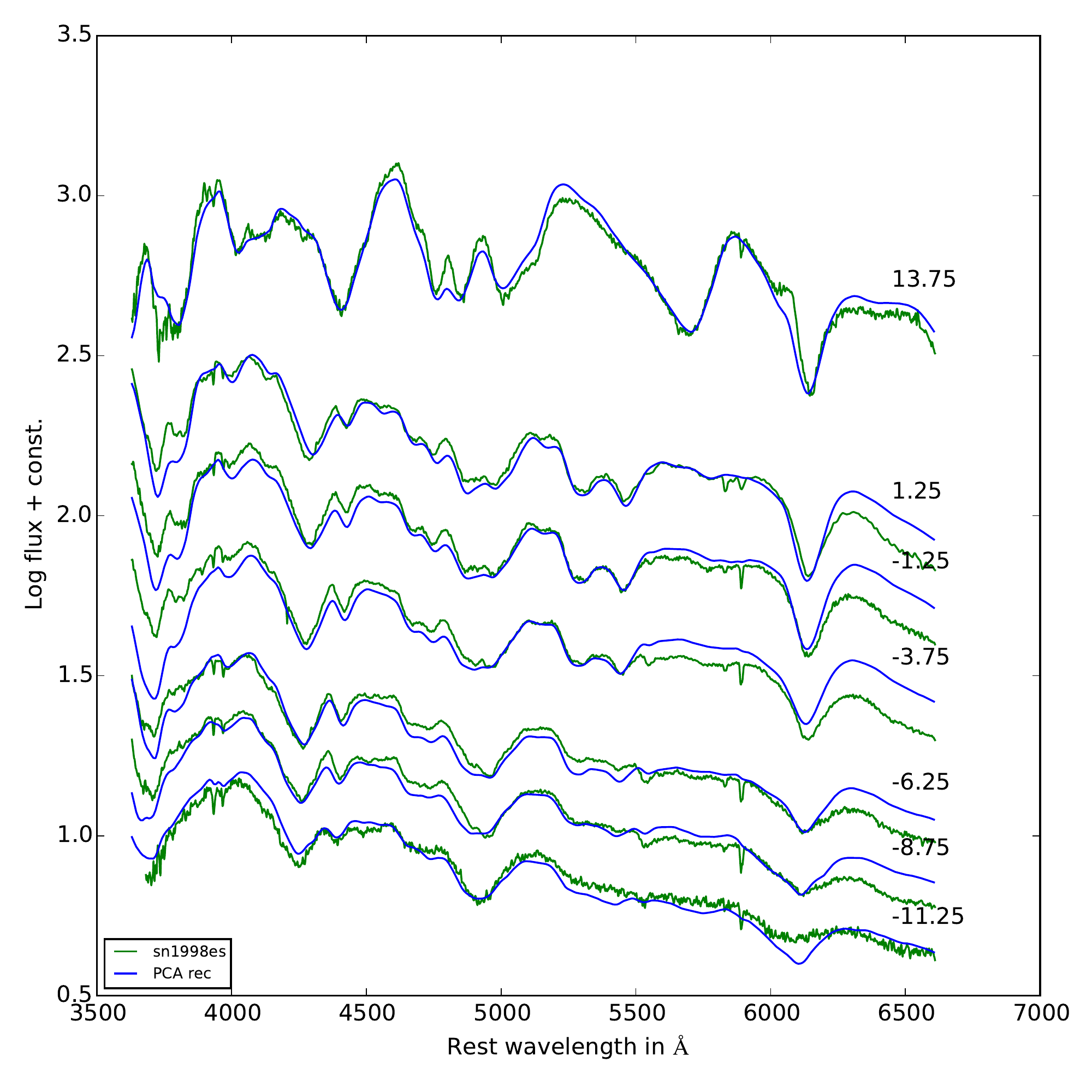}}
\caption{Observed spectra of SN 1998es (in green) with the
  reconstructed spectra (in blue) overplotted. The data are from
  \protect\cite{2012AJ....143..126B} and
  \protect\cite{2012MNRAS.425.1789S}. }
\label{fig:spectra_sn1998es}
\end{figure*}

\begin{figure*}
\centering
\subfloat{\includegraphics[width=1.2\columnwidth]{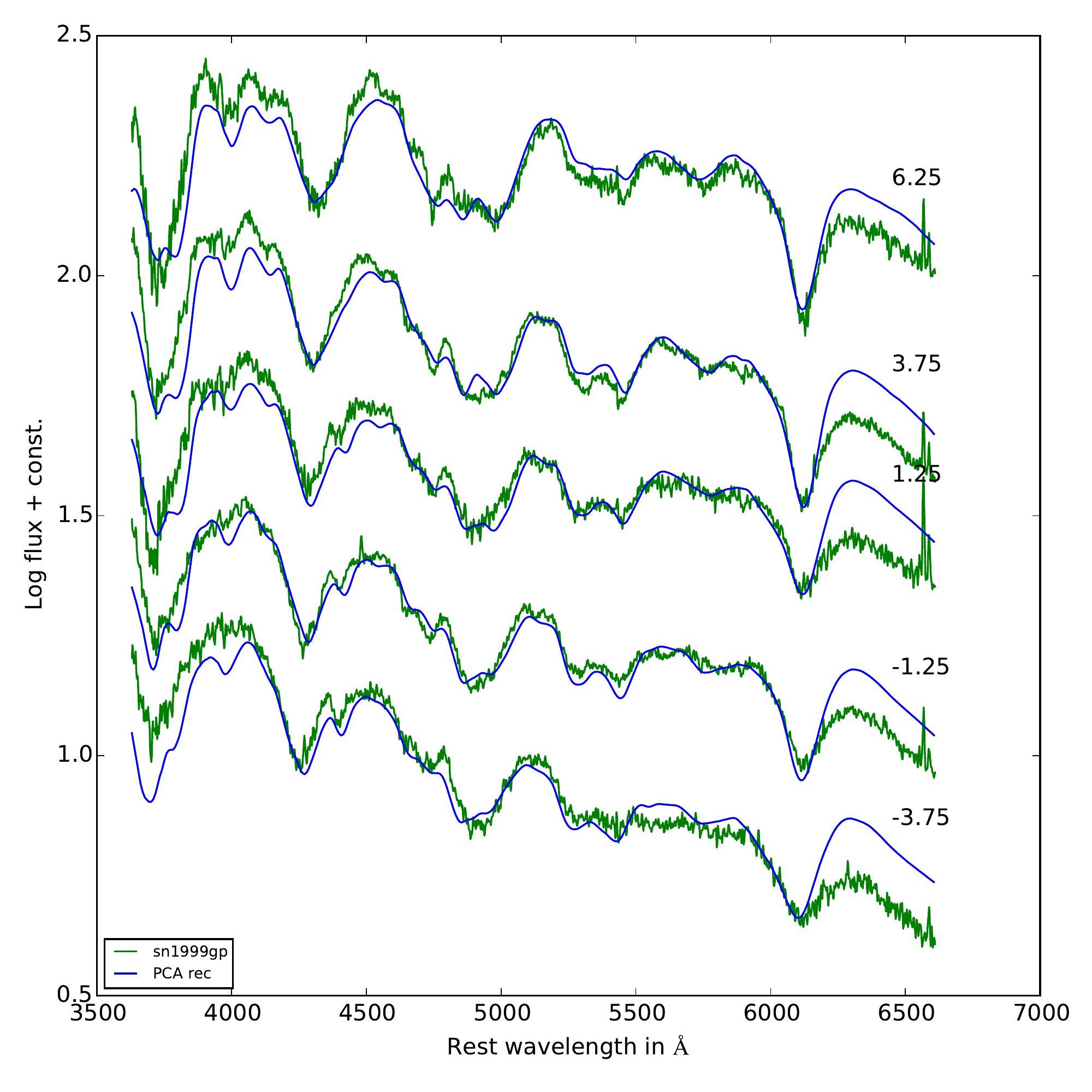}}
\caption{Observed spectra of SN 1999gp (in green) with the
  reconstructed spectra (in blue) overplotted. The data are from
  \protect\cite{2012AJ....143..126B}. }
\label{fig:spectra_sn1999gp}
\end{figure*}

This also holds for 91T-like events, as shown in
Figs. \ref{fig:spectra_sn1998es} and \ref{fig:spectra_sn1999gp}. For
the well observed (16 spectra) SN 1998es in NGC 632, a typical member
of this group, besides again a slight missmatch of color, the
reconstruction is good, and it is almost perfect for the late epochs,
and the same holds for SN 1999gp (6 spectra).

\begin{figure*}
\centering
\subfloat{\includegraphics[width=1.2\columnwidth]{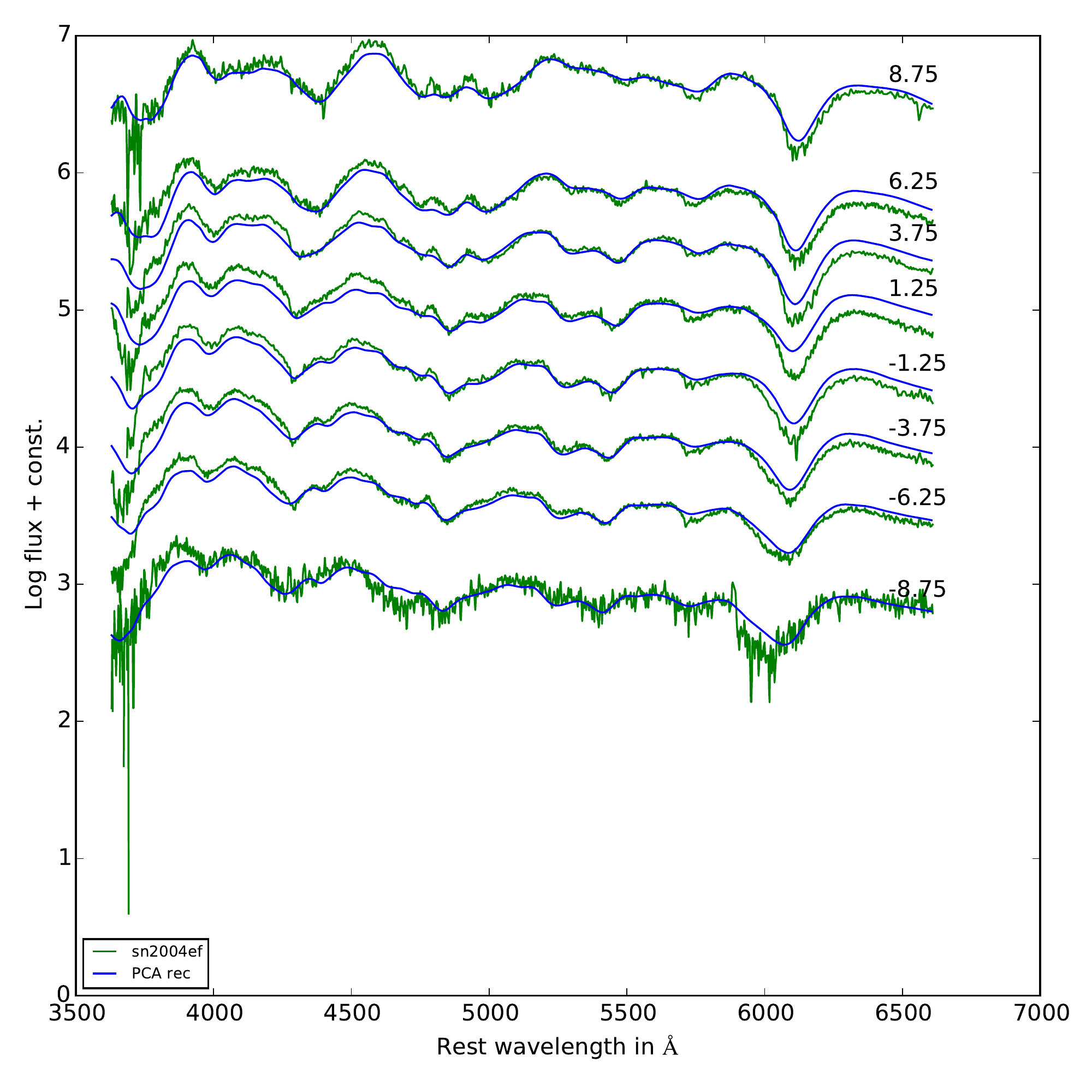}}
\caption{Observed spectra of SN 2004ef (in green) with the
  reconstructed spectra (in blue) overplotted. The data are from
  \protect\cite{2012AJ....143..126B},
  \protect\cite{2012MNRAS.425.1789S}, and
  \protect\cite{2013ApJ...773...53F}. }
\label{fig:spectra_sn2004ef}
\end{figure*}

\begin{figure*}
\centering
\subfloat{\includegraphics[width=1.2\columnwidth]{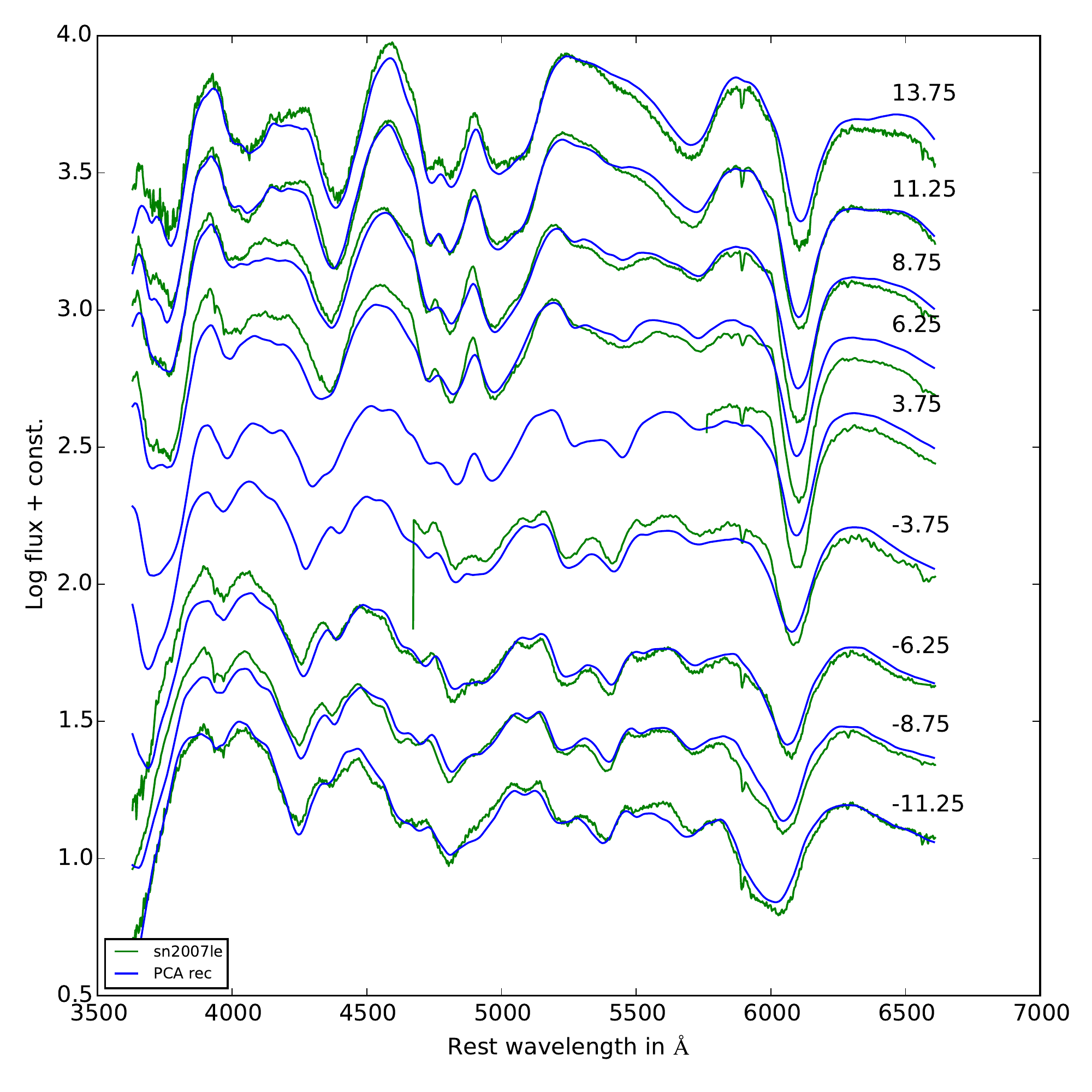}}
\caption{Observed spectra of SN 2007le (in green) with the
  reconstructed spectra (in blue) overplotted. The data are from
  \protect\cite{2012AJ....143..126B},
  \protect\cite{2012MNRAS.425.1789S}, and
  \protect\cite{2013ApJ...773...53F}. }
\label{fig:spectra_sn2007le}
\end{figure*}

SN 2004ef (21 spectra) and SN 2007le (32 spectra), shown in
Figs. \ref{fig:spectra_sn2004ef} and \ref{fig:spectra_sn2007le}, are
  members of the group of supernovae with high photospheric
  \SiII\ velocities. They populate a rather large part of the PCA
  space, indicating significant spectroscopic diversity. Given this
  fact, the reconstruction works surprisingly well. Again, we find
  some color missmatch around maximum light, but otherwise the fits
  are good in both cases. Note that even spectra with incomplete wave
  band coverage (as for SN 2007le) are reconstructed rather well.

\begin{figure*}
\centering
\subfloat{\includegraphics[width=1.2\columnwidth]{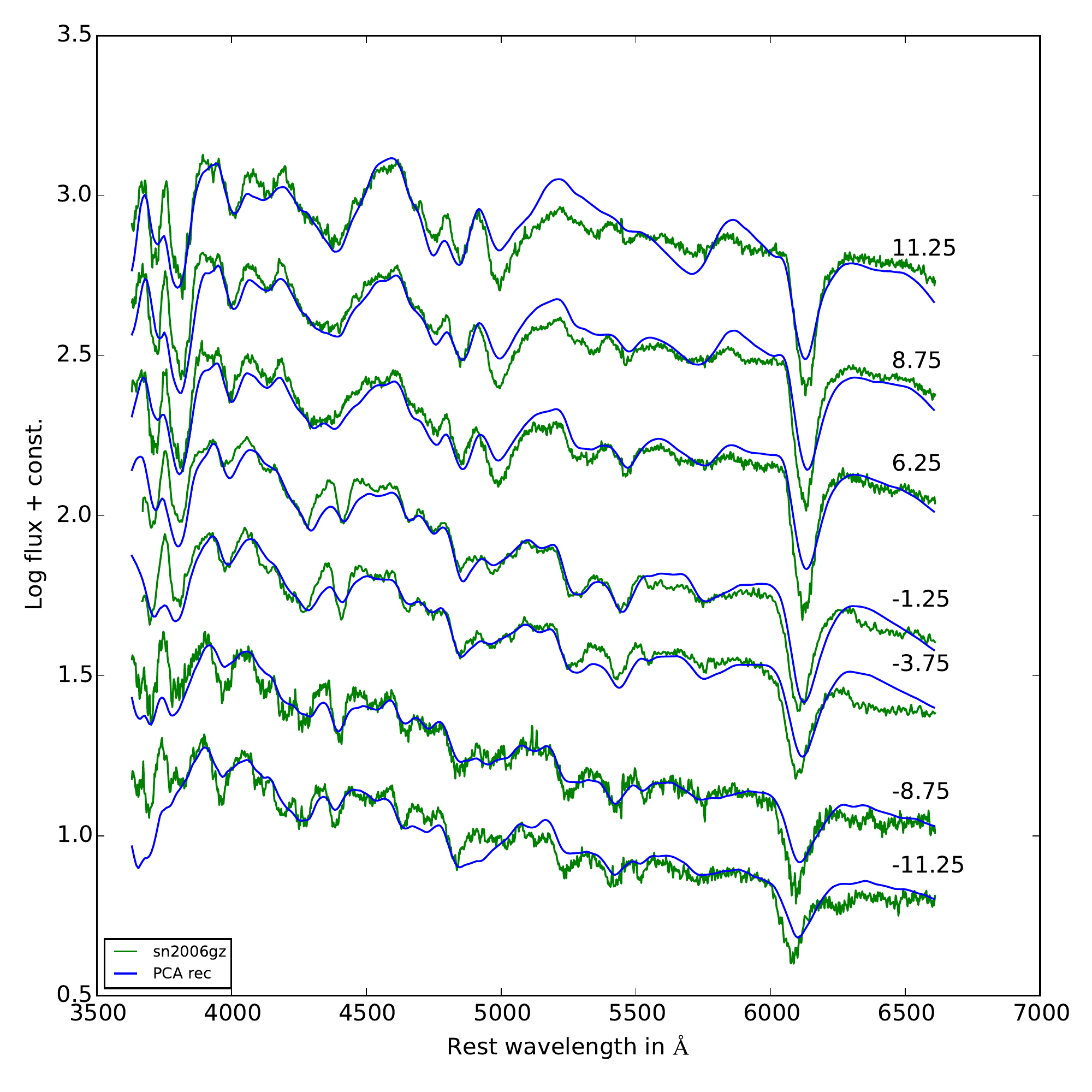}}
\caption{Observed spectra of SN 2006gz (in green) with the
  reconstructed spectra (in blue) overplotted. The data are from
  \protect\cite{2012AJ....143..126B}, and SUSPECT \citep{2007ApJ...669L..17H}. }
\label{fig:spectra_sn2006gz}
\end{figure*}

Finally, we show the reconstruction for SN 2006gz (24 spectra), a
peculiar, very luminous carbon-rich supernova, considered to be the
prototype of so-called ``super-Chandrasekhar mass'' explosions (Fig
\ref{fig:spectra_sn2006gz}). As in the cases of 91bg-like and 91T-like
supernovae there are very few members in the training
set. Nevertheless, the reconstruction works surprisingly well,
especially after peak.

In conclusion, the PCA reconstruction of multi-epoch spectra of SNe Ia
works reasonably fine even in cases where the limited number of objects
in our training set might have indicated a problem. Also the data
quality is not an issue and is taken care of by the method. Therefore
we are confident that we can use the metric space constructed for a
comparison with synthetic spectra of models.

\section{PCA reconstruction of model spectra and their
comparison with real supernovae}
\label{sec:appendix_b}

In this Appendix we show spectra of several of the models used in this
paper and compare them with their PCA reconstruction. Such a
comparison is necessary since the model spectra are not part of the
training set and, therefore, it is not obvious that the reconstructed
synthetic spectra do resemble the 'original' ones, in particular
given the fact that our PCA is based on the derivatives of the
spectra. We will do the comparison for one or two typical
realizations of each class of models discussed in the main text. As in
Appendix A, in all figures the logarithm of flux over wavelength is
plotted, and each spectrum is labeled with the time relative to B-band
maximum. Time is progressing from bottom to top. Computed model spectra
are in green and their reconstructed counterparts are in blue. All
model spectra are viewing angle averaged. 
Note that we plot the observed spectra rather than their PCA
reconstruction which makes the comparison even more demanding.
As in Appendix A, the epochs are binned and we use 5 PCs for the
reconstruction.

\begin{figure*}
\centering
\subfloat{\includegraphics[width=1.2\columnwidth]{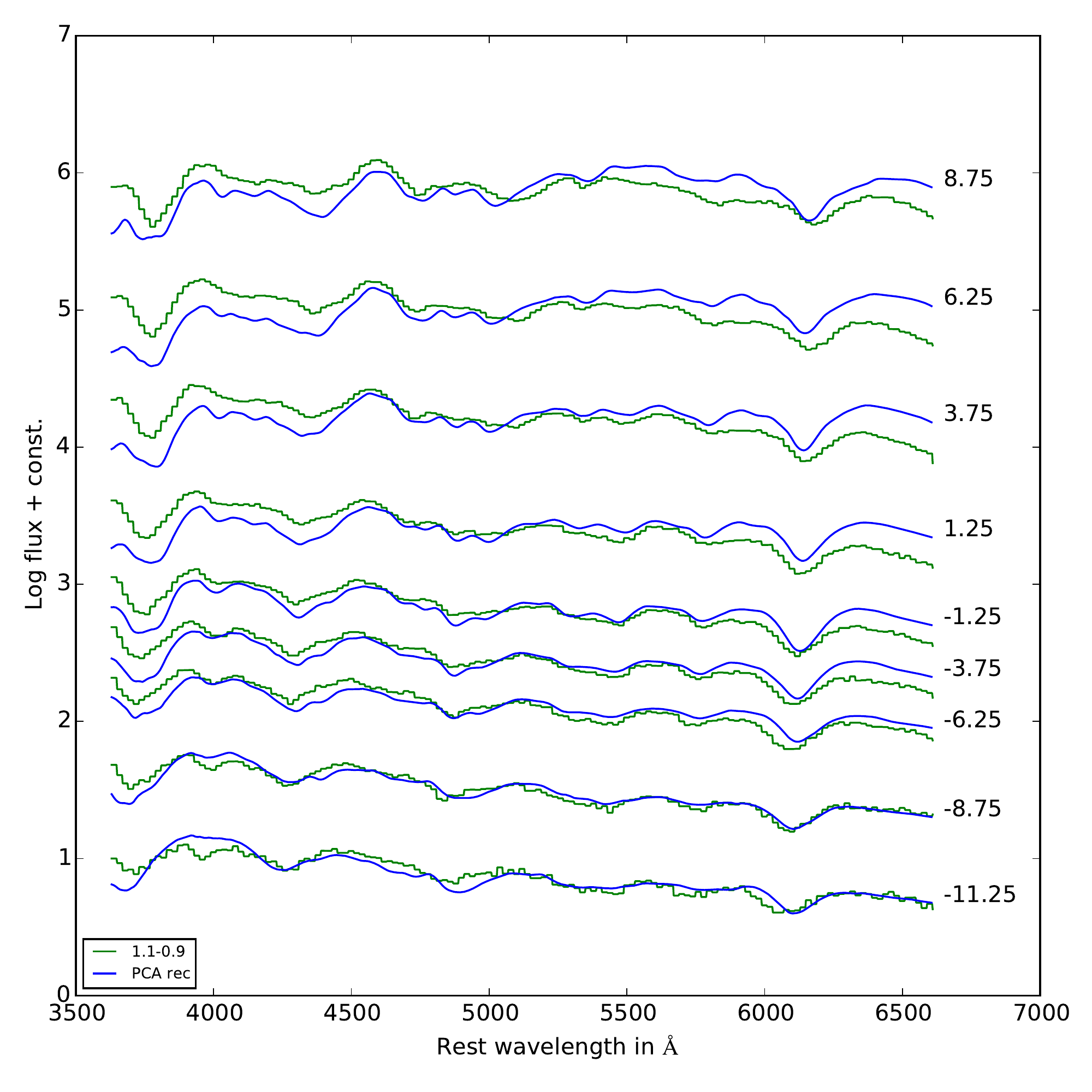}}
\caption{Synthetic spectra of the merger model (1.1 M$_\odot$ + 0.9
  M$_\odot$) overplotted with the reconstructed spectra. The model
  spectra are in green and the reconstructions are in blue. The
  spectra are labeled with time relative to B-band maximum. }
\label{fig:spectra_1.1-0.9}
\end{figure*}

\begin{figure*}
\centering
\subfloat{\includegraphics[width=1.2\columnwidth]{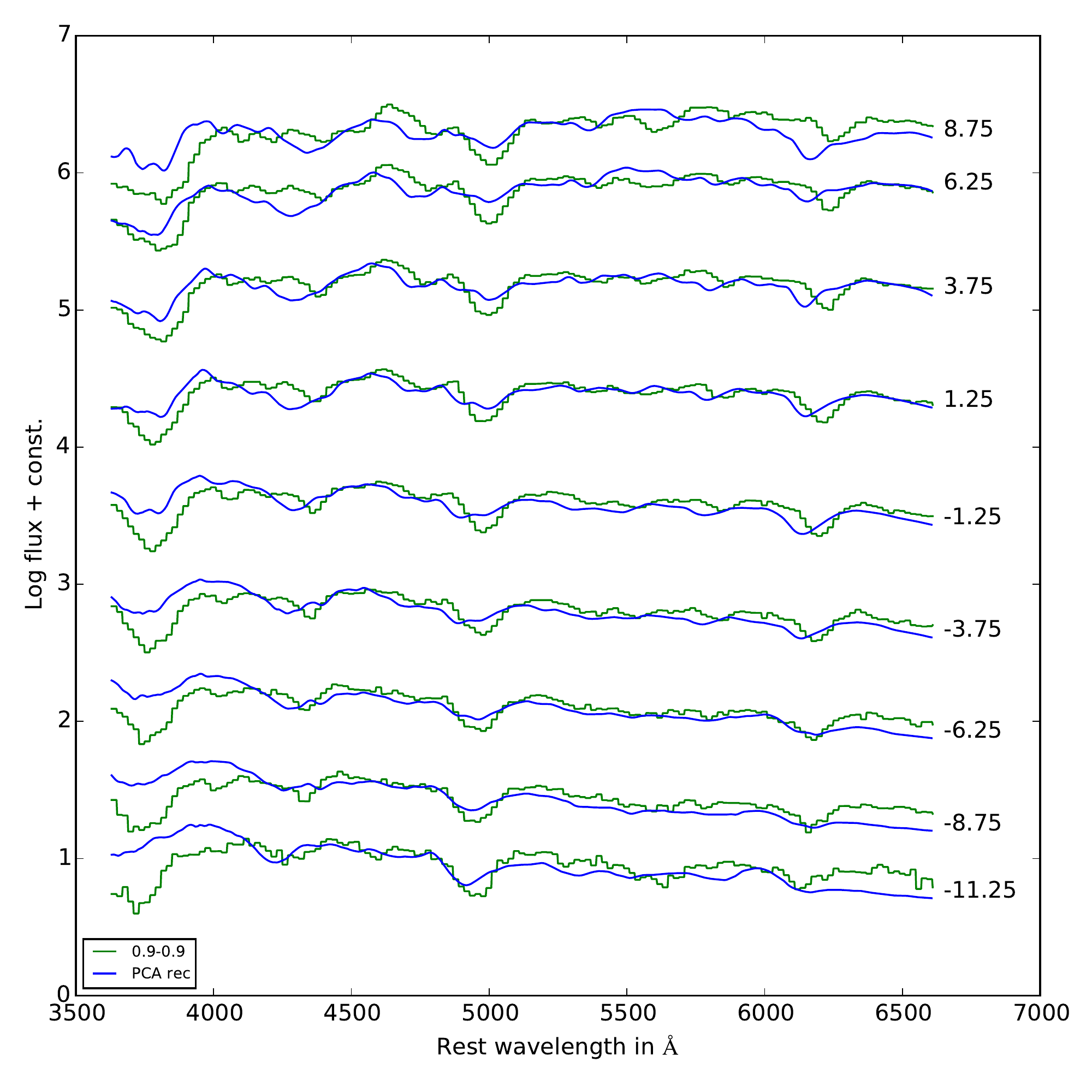}}
\caption{Same as \ref{fig:spectra_1.1-0.9}, but for the merger
  model (0.9 M$_\odot$ + 0.9 M$_\odot$). }
\label{fig:spectra_0.9-0.9}
\end{figure*}

We start with the violent merger models.
Fig.~\ref{fig:spectra_1.1-0.9} shows the spectra predicted for the
merger of a 1.1 M$_\odot$ with a 0.9 M$_\odot$ white dwarf.  This
model was used for a comparison with the normal SN 2011fe by R\"opke
et al. (2012). The agreement between computed and reconstructed
spectra is good for early epochs, but gets worse at late times where
also the color offset is more pronounced. The second merger model we
present is for two equal mass (0.9 M$_\odot$) white dwarfs, a model that
is known to show similarities with 91bg-like events
(Fig.~\ref{fig:spectra_0.9-0.9}) and has been used in this context by
Pakmor et al. (2010). Here, the reconstructions are never really good,
getting worse for the later spectra. The quality of the reconstruction
of 91bg-like is not as good as other more populated subtypes of SNe.
This is the likely reason for this discrepancy. Note that also the
quality of the reproduction of observed 91bg-like supernovae was not
great (see Appendix A).

\begin{figure*}
\centering
\subfloat{\includegraphics[width=1.2\columnwidth]{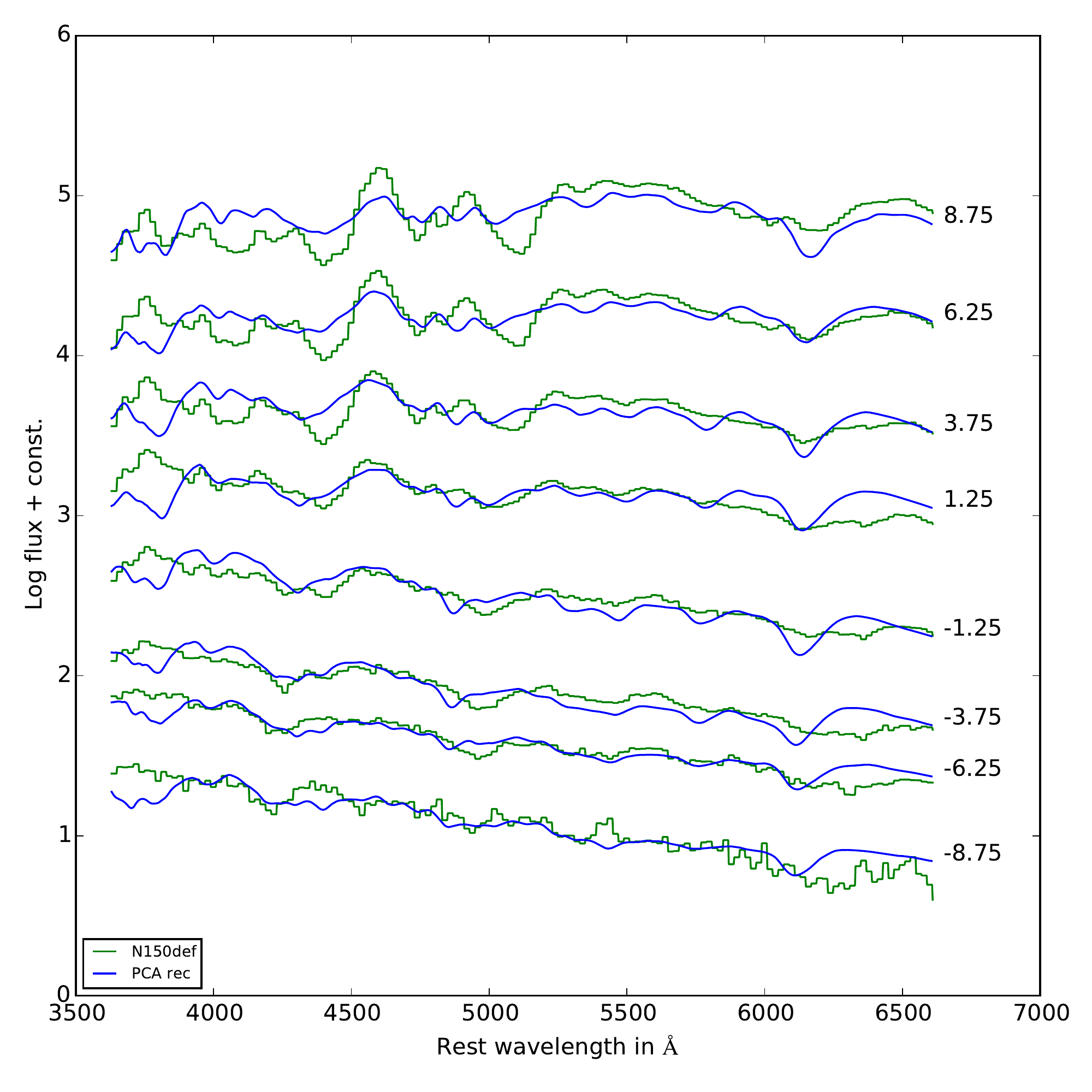}}
\caption{Same as \ref{fig:spectra_1.1-0.9}, but for the pure
  deflagation model N150def. }
\label{fig:spectra_N150def}
\end{figure*}

Pure deflagration models were suggested to be good candidates for
(02cx-like) peculiar supernovae and, in fact, in Figs. 3 and 4 we find
them in the matching places of the PCA space. However, since again
there are only a few objects of this group in our training set we do
not expect to find very good fits. On the other hand, some pure
deflagration models have $^{56}$Ni masses in the range of normal SNe
Ia. So, in principle, these qualify for the fainter normal ones as
well. Fig. \ref {fig:spectra_N150def} shows model N150def, a typical
example of relatively bright pure deflagation which disrupts the
Chandrasekhar-mass white dwarf producing about 0.4 M$_\odot$ of
$^{56}$Ni. It is obvious that in particular past maximum the
reconstructed spectra do not match the computed ones. For instance,
the PCA reconstruction shows a \SiII\ feature at 6100~\AA ~which is not
present in the model spectra. At the same time, features in the model
spectra at shorter wave length, which are not present in the data, are
weakened or even ignored in the reconstructions.

\begin{figure*}
\centering
\subfloat{\includegraphics[width=1.2\columnwidth]{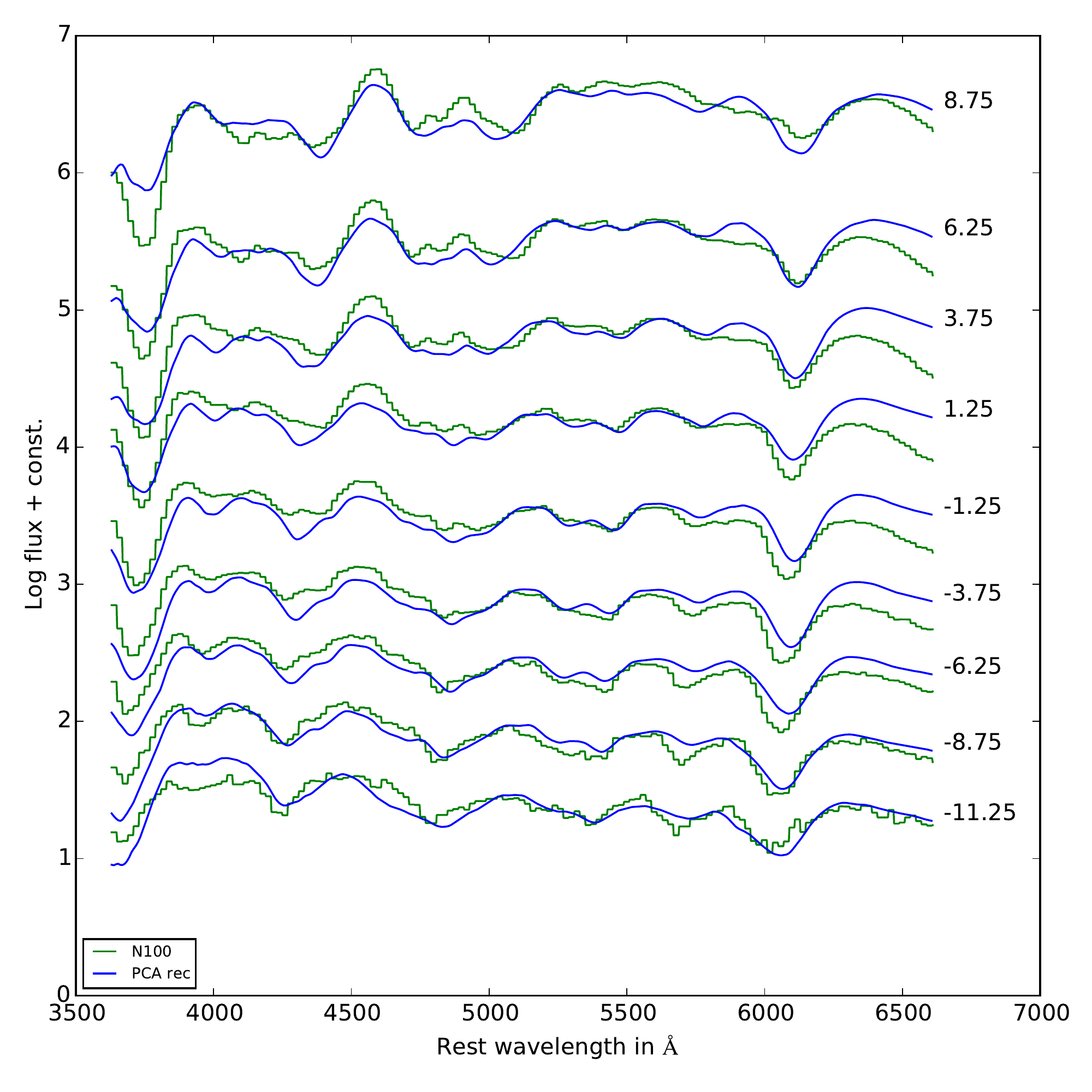}}
\caption{Same as \ref{fig:spectra_1.1-0.9}, but for the
  delayed-detonation model N100. }
\label{fig:spectra_N100}
\end{figure*}

\begin{figure*}
\centering
\subfloat{\includegraphics[width=1.2\columnwidth]{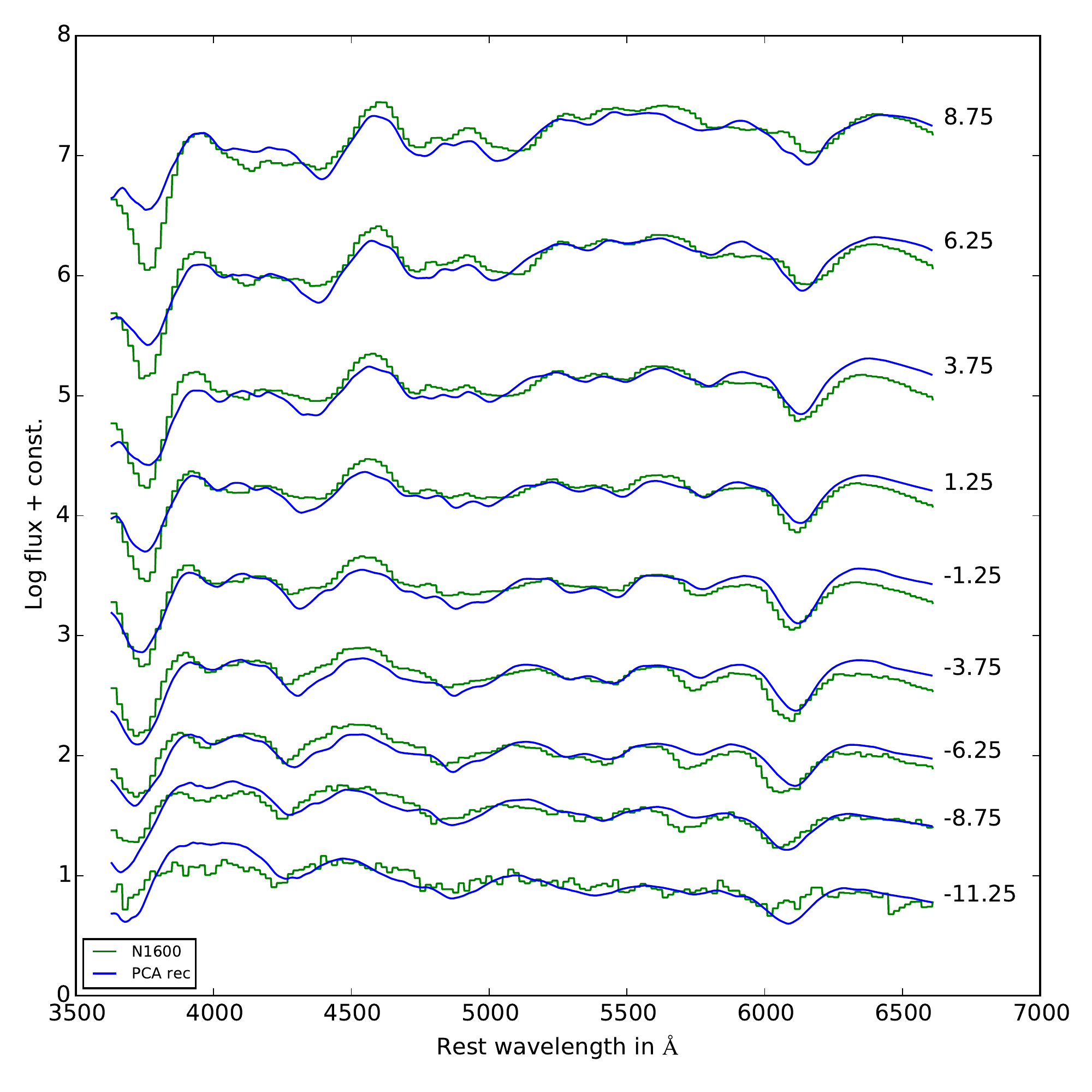}}
\caption{Same as \ref{fig:spectra_1.1-0.9}, but for the
  delayed-detonation model N1600. }
\label{fig:spectra_N1600}
\end{figure*}

Models in which the deflagation wave changes into a detonation during
the explosion are also candidates for normal SNe Ia because they can,
in principle, explain the range of observed luminosities. The model we
show in Fig.~\ref{fig:spectra_N100} (``N100'') is a typical example of
this class and was also used for a comparison with SN 2011fe by Roepke
et al. (2012).  The reconstruction works significantly better in this
case, in particular for early epochs. This is not too surprising since
at these epochs the model resembles major spectral features of normal
SN Ia rather well (see also Fig. B9).  Similarly, for model ``N1600''
which, according to Figures 3 and 4, should be a good model for some
of the observed explosions, also the reconstruction works properly, as
shown in Fig.~\ref{fig:spectra_N1600}.

\begin{figure*}
\centering
\subfloat{\includegraphics[width=1.2\columnwidth]{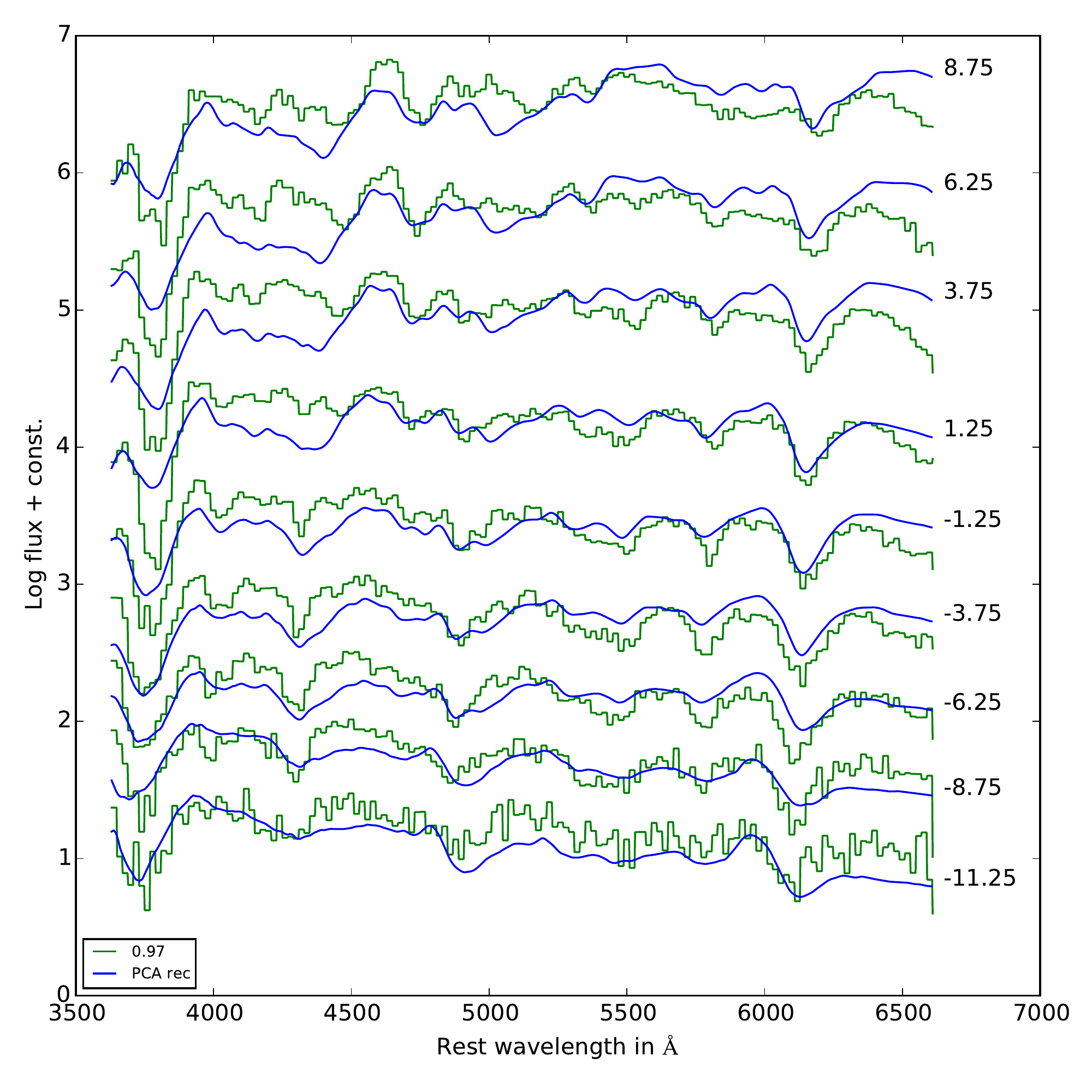}}
\caption{Same as \ref{fig:spectra_1.1-0.9}, but for the
  sub-Chandrasekhar mass model of 0.97 M$_\odot$. }
\label{fig:spectra_0.97}
\end{figure*}

\begin{figure*}
\centering
\subfloat{\includegraphics[width=1.2\columnwidth]{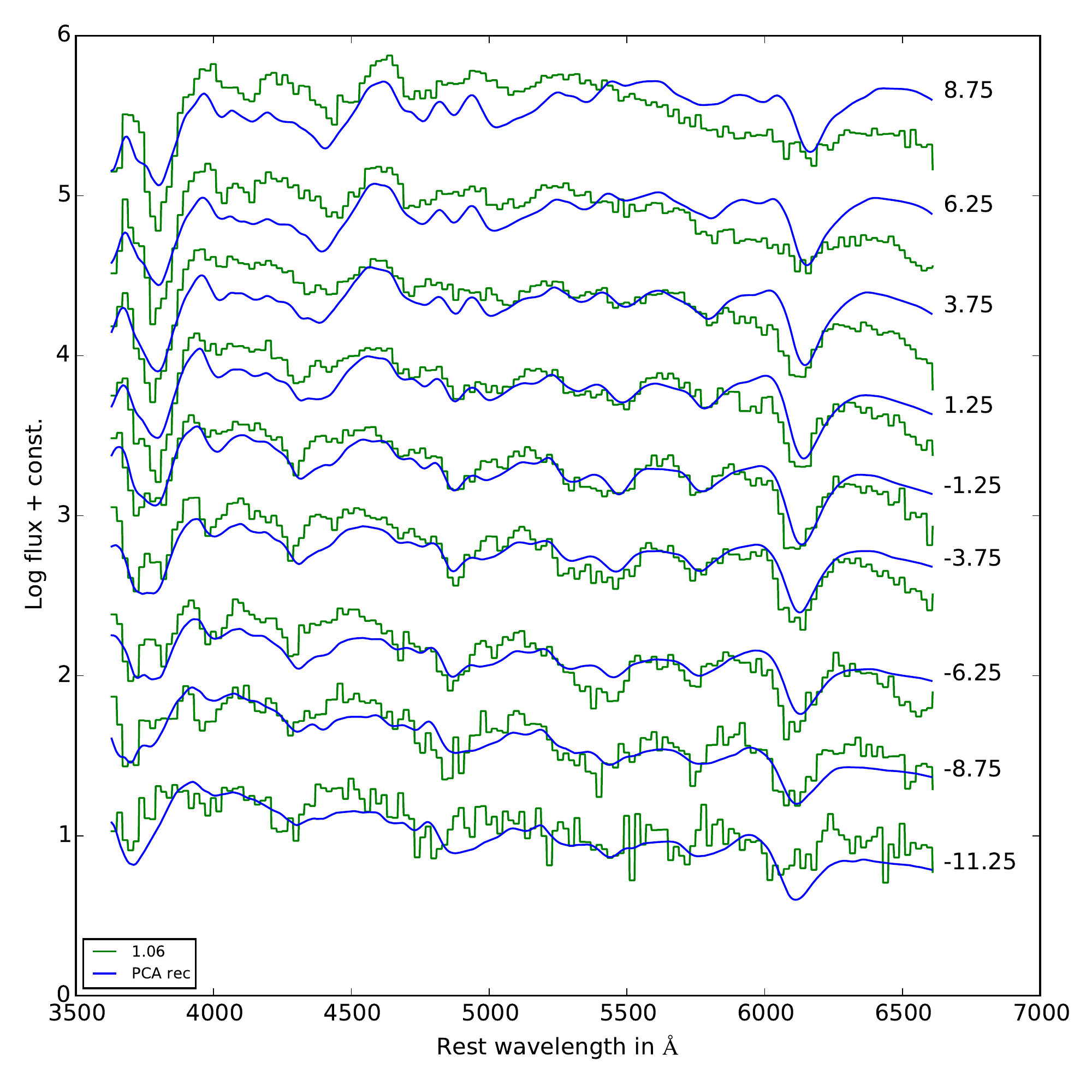}}
\caption{SyntSame as \ref{fig:spectra_1.1-0.9}, but for the
  sub-Chandrasekhar mass model of 1.06 M$_\odot$. }
\label{fig:spectra_1.06}
\end{figure*}

Finally, we compare two of the sub-Chandrasekhar mass (pure
detonation) models with their reconstruction. We have chosen the
models with 0.97 M$_\odot$ (Fig.~\ref{fig:spectra_0.97} ) and 1.06
M$_\odot$ (Fig.~\ref{fig:spectra_1.06}), respectively, since they
produce Ni-masses in the ballpark of normal SNe Ia. Not unexpectedly,
the reconstruction is very good in both cases.  There are small
discrepancies for the spectra two to three weeks past maximum, but in
general, both, spectral features and their evolution with time are
reproduced.

In conclusion, we have demonstrated that it is possible to represent
the time sequences of model spectra by means of their principle
components computed in the PCA space of the data. This works as long
as the synthetic spectra bear some similarities with real supernova
spectra but looses reliability if this is not the case or if the
models are close to the edges of the PCA space populated by the data.
The first problem becomes obvious when comparing synthetic spectra
with their reconstructed counterparts at later times (t$>$ 10 days
past maximum) where our radiative-transfer modeling becomes less
reliable.

\begin{figure*}
\centering
\subfloat{\includegraphics[width=1.2\columnwidth]{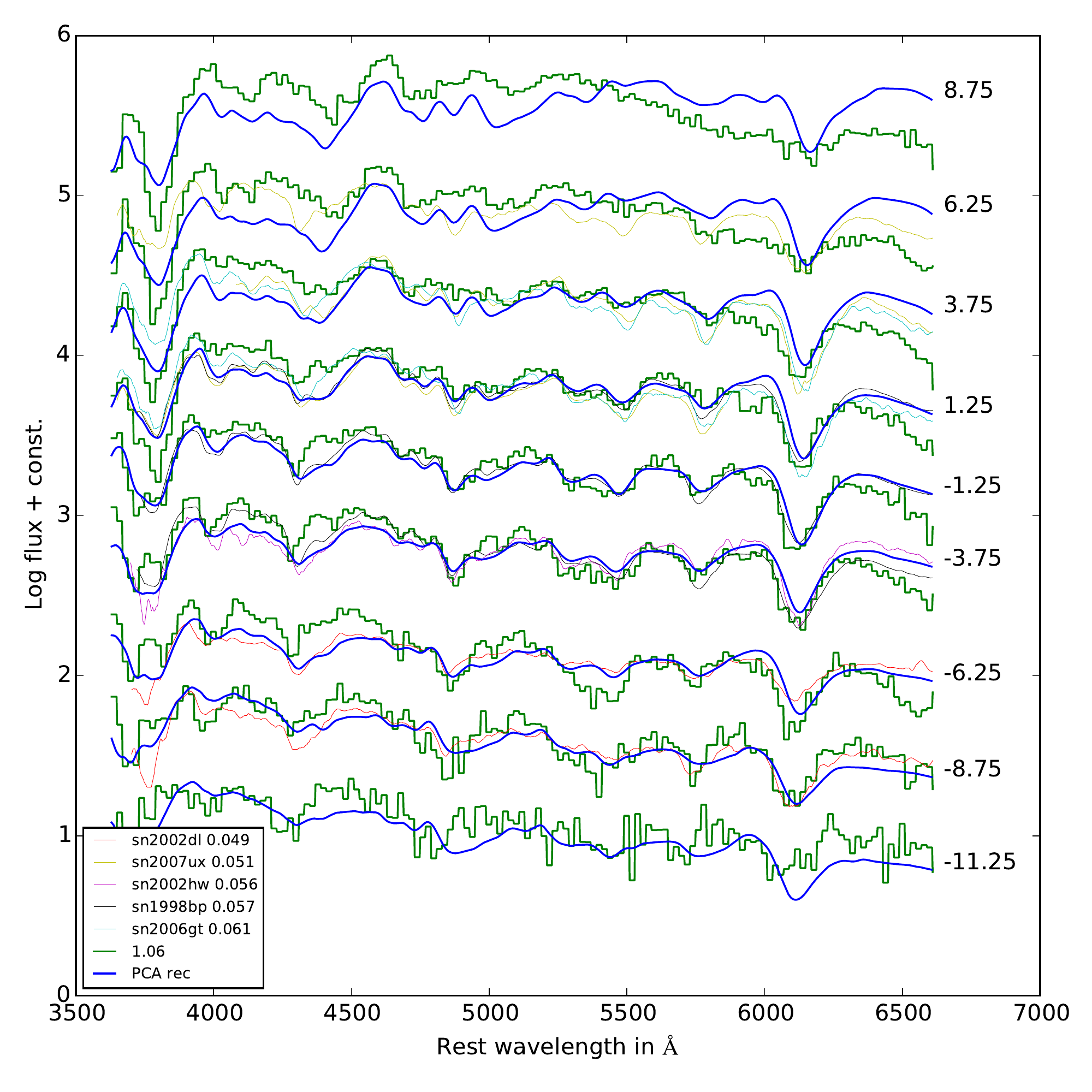}}
\caption{Synthetic spectra of the sub-Chandrasekhar mass
  model of 1.06 M$_\odot$ (in green) and their reconstruction
  (in blue) overplotted with spectra of nearby (in PCA space) observed
  supernovae. These supernovae a listed in the box in the lower left
  corner. The distance between the model and the SN in the PCA space
 is listed beside the name. }
\label{fig:spectra_next_to_1.06}
\end{figure*}

\begin{figure*}
\centering
\subfloat{\includegraphics[width=1.2\columnwidth]{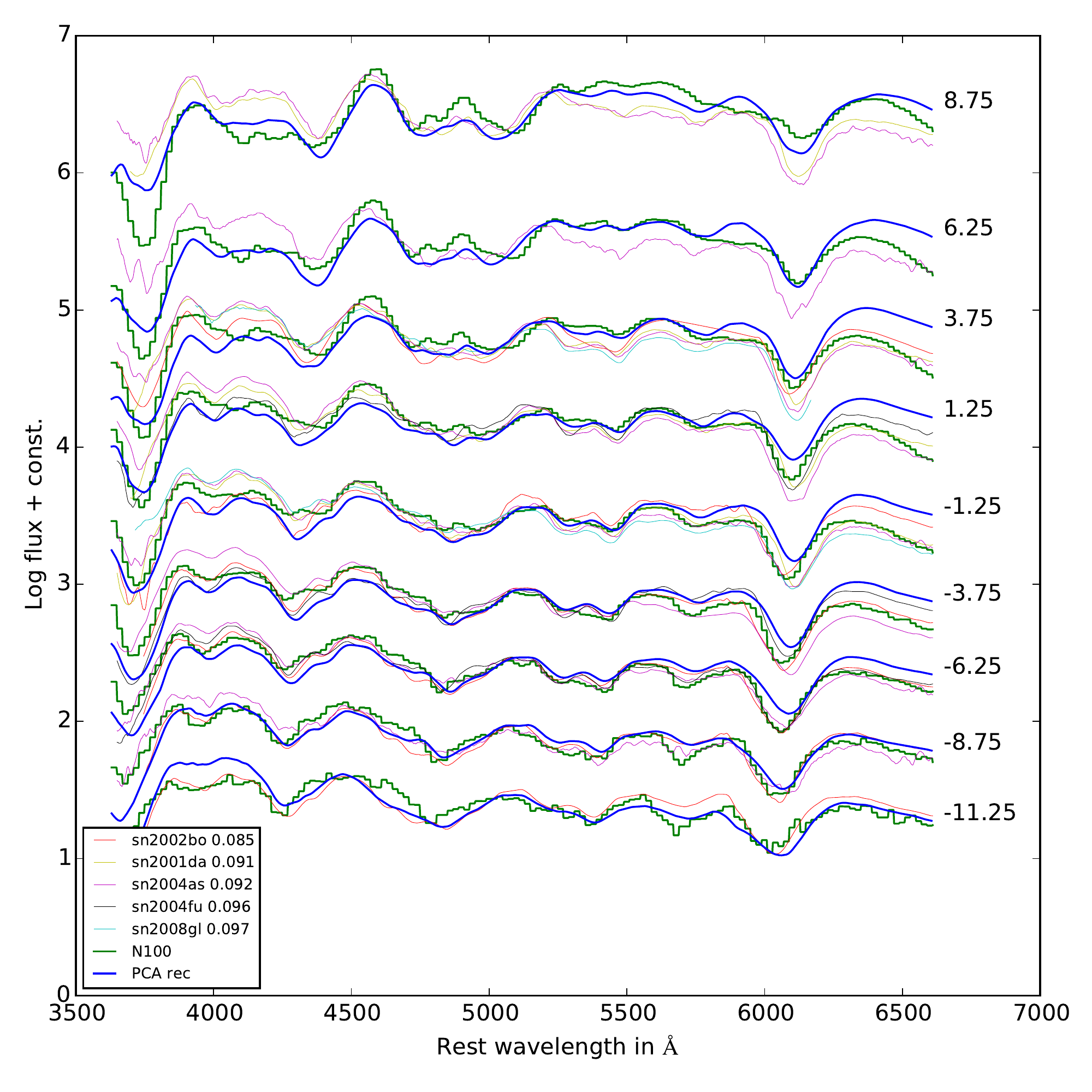}}
\caption{Same as \ref{fig:spectra_next_to_1.06}, but for the
  delayed-detonation model N100. }
\label{fig:spectra_next_to_N100}
\end{figure*}

\begin{figure*}
\centering
\subfloat{\includegraphics[width=1.2\columnwidth]{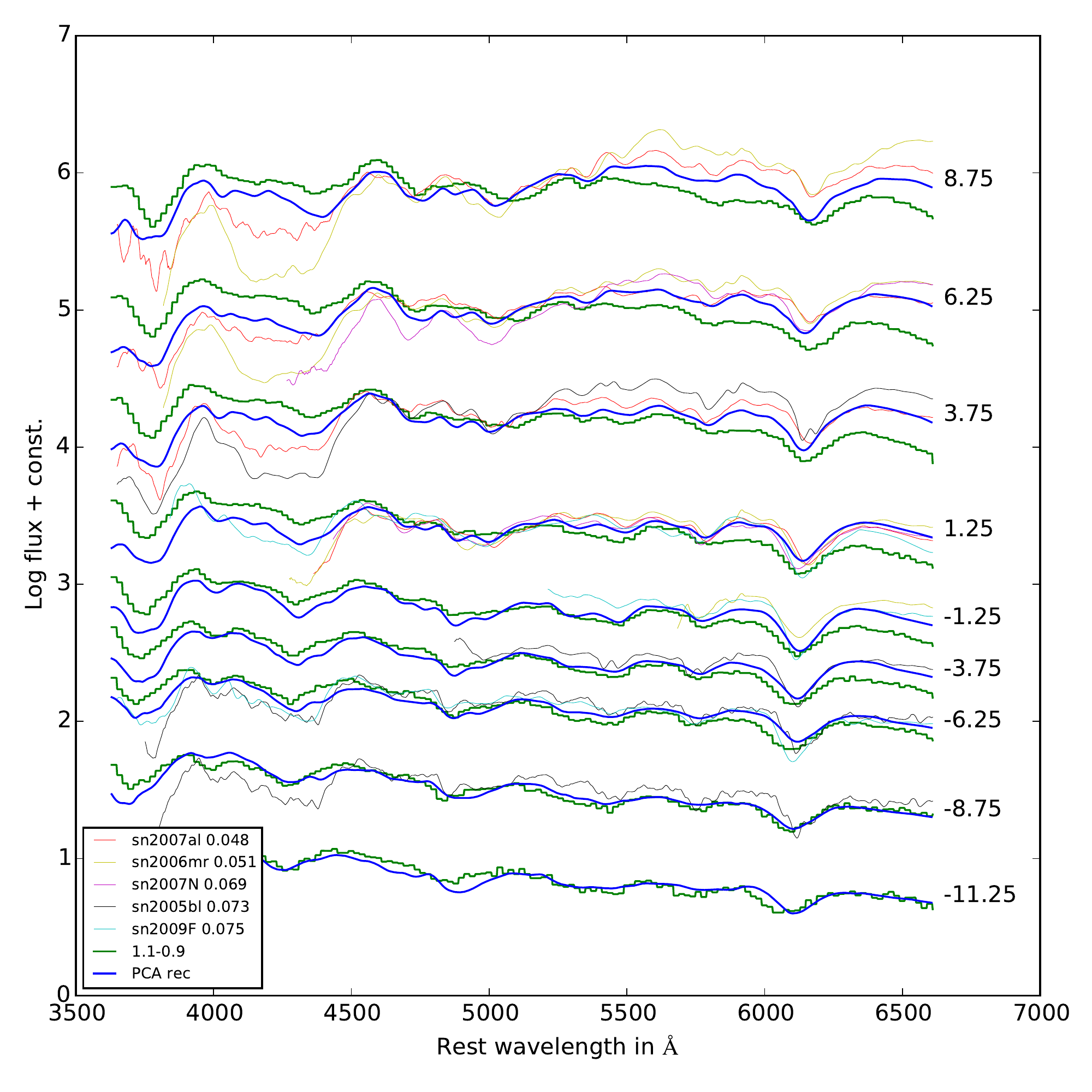}}
\caption{Same as \ref{fig:spectra_next_to_1.06}, but for the
  violent-merger model (1.1 M$_\odot$ + 0.9 M$_\odot$). }
\label{fig:spectra_next_to_1.1-0.9}
\end{figure*}

Of course, a fair question is: How do model spectra compare with their
nearest neighbors in PCA space? Ideally, one would hope that if in PCA
space there are one or several real supernova nearby to a model their
spectra should agree well. Since by means of PCA we have constructed a
metric space such a comparison can be done easily, and we show the
results in Figs.~\ref{fig:spectra_next_to_1.06} through
~\ref{fig:spectra_next_to_1.1-0.9}.

In Fig.~\ref{fig:spectra_next_to_1.06} we play the game for the
sub-Chandrasekhar mass model of Figure B6. We plot the five nearest
(in PCA space) supernovae listed in the box in the left corner of the
figure. The distances between the SNe and the model is listed in the
legend beside their names. The model spectra are in green and their
PCA reconstruction is in blue. The agreement is amazing, in particular
for the pre-maximum spectra. Note that a close match to this model is
also SN 2002dl, a 91bg-like supernova.  Similarly, we find rather good
agreement between the delayed-detonation model 'N100' (Fig. B4) and
its reconstruction and some of its neighbors in PCA space
(Fig.~\ref{fig:spectra_next_to_N100}). This time, however, these
neighbors are all normal or High Velocity SNe Ia which agrees with
what we see in Figures 3 and 4. Note, however, that their distance to
the model is quite large, indicating significant differences coming
mainly from PC 4 and 5 which are not plotted. As our last example we
show the merger model of Figure B1 in comparison with its nearest
neighbors (Fig.~\ref{fig:spectra_next_to_1.1-0.9}). This time the fits
are not quite as good, in particular at late times where also the
reconstruction gets worse.  This might be due to the approximations of
the radiation transport procedure as all models differ from the
corresponding reconstructions in a similar way.

All in all we find that our method passes also this final test rather
well. It is not perfect and has to be applied with care, but it
opens the possibility to compare models with observations in a more
quantitative way and for large samples.

\bibliography{biblio}


\end{document}